\documentclass[12pt,epsf]{article}

 \usepackage{epsfig}
 \usepackage{times}
 
 \usepackage{amsmath}
 \usepackage{amssymb}

\newcommand{\be}{\begin{equation}}
\newcommand{\ee}{\end{equation}}
\def\({\left (}
\def\){\right )}
\def\[{\left [}
\def\[{\right ]}

\begin{document}
\begin{titlepage}
\bigskip
\rightline{hep-th/0706.3677}
\rightline
\bigskip\bigskip\bigskip\bigskip
\centerline {\Large \bf {Bubbles Unbound II: AdS and the Single Bubble}}
\bigskip\bigskip
\bigskip\bigskip

\centerline{\large Keith Copsey}
\bigskip\bigskip
\centerline{\em Department of Physics, Broida Hall, UCSB, Santa Barbara, CA 93106}
\centerline{\em keith@physics.ucsb.edu}
\bigskip\bigskip

\begin{abstract}
I present further analytic time symmetric initial data for five dimensions describing ``bubbles of nothing'' which have no  Kaluza-Klein circle asymptotically.   The new solutions consist of a large family of single bubbles in both asymptotically flat and asymptotically AdS space.  I also numerically construct gravitational solitons in AdS where the usual boundary conditions are modified geometrically but not topologically.  Furthermore I point out there are both regular AdS bubbles and topologically trivial metrics in asymptotically global AdS dual to matter which violates all the usual energy conditions, both classical and quantum.   I inquire as to the existence of a dual gravitational instability.

\end{abstract}
\end{titlepage}

\baselineskip=16pt
\setcounter{equation}{0}

\section{Introduction}

Kaluza-Klein ``bubbles of nothing'' were introduced a quarter of a century ago by Witten \cite{WittenBubbles} as an instability in the Kaluza-Klein (KK) vacuum.  By performing an analytic continuation on a Schwarzschild black hole he was able to find an instanton which describes the nucleation of a ``bubble'' where the Kaluza-Klein circle smoothly pinches off in the interior of the spacetime, resulting in a minimal two sphere.  Once produced the bubble accelerates out to null infinity, ``eating'' up the spacetime.  The production of these bubbles is fortunately forbidden in a theory with fundamental fermions and supersymmetric boundary conditions.   At the point where the circle pinches off (the end of the ``cigar'') the fermions are, by definition, antiperiodic.  Since the cigar is a simply connected manifold with a single spin structure a KK bubble requires antiperiodic boundary conditions for the fermions at infinity.  These boundary conditions are, however, inconsistent with supersymmetry.  In the intervening years since their introduction bubbles have been useful in a wide variety of applications in time dependent spacetimes and black hole physics (see e.g. \cite{Elvangetal}  and \cite{HorowitzMaeda} and references therein).

In \cite{CopseyBubbles1} time symmetric initial data describing purely gravitational bubbles in five dimensional asymptotically flat space were found by considering a spatial metric motivated by the black ring solutions \cite{EmpReall}.   These bubbles locally look like KK bubbles, but are asymptotically globally flat.  The particular solutions consist of two minimal two spheres which touch at a single point.  These solutions have two free parameters which may be taken to be the sizes of the two bubbles or, equivalently, the mass of the solution and the size of the larger bubble.  If the two bubbles are of comparable size both initially collapse  but if one is significantly larger than the other the larger bubble begins expanding while the smaller bubble collapses.  If this expansion continues for any significant period of time, the spacetime far away from the initial disturbance is radically altered.  Hence these solutions represent a new possible instability of asymptotically flat space.   Note the simple argument given by Witten to rule out Kaluza-Klein bubbles with supersymmetric boundary conditions does not apply here; the circle associated with the bubble is absent by the time one reaches infinity.

It is natural to wonder whether there are single asymptotically flat bubbles.  Further, in view of the AdS-CFT correspondence, one would like to be able to obtain such solutions in an asymptotically AdS context.  The present work describes such solutions.   Building on an ansatz based on static charged bubbles found by Ross \cite{RossBubbles}, which in turn were motivated by the solutions of \cite{CveticBH}, I construct a large class of time symmetric initial data involving only the metric and possibly a cosmological constant.    The solutions may also be viewed as a large generalization of the single asymptotically flat solution found by LeBrun  \cite{LeBrun} some time ago.

I begin by constructing of these solutions and then describe their geometric properties.    I then make some comments regarding the dynamics of these bubbles and in particular note that there are many expanding bubbles in both asymptotically flat and AdS space.  Furthermore I show that if one is willing to consider AdS boundary conditions where the asymptotic sphere is squashed instead of round there exist a series of static regular bubbles.   In the AdS context I calculate the dual stress boundary tensor and demonstrate the new solutions include both bubbles and topologically trivial metrics are dual to matter which violates all standard classical and quantum energy bounds.  This strongly suggests the existence of an instability in the dual field theory.  Finally, I begin an inquiry as to the existence of a dual gravitational instability.

 \setcounter{equation}{0}
\section{New Solutions}
\subsection{Original Form}

Motivated by Ross's solutions  \cite{RossBubbles} consider a spatial metric containing a
\newline
 squashed three sphere
\be \label{ansatz}
ds^2 = \frac{dr^2}{W(r)} + \alpha(r) \Big( {\sigma_1}^2 + {\sigma_2}^2 \Big) + \beta(r) {\sigma_3}^2 
\ee
where $\sigma_i$ are the left-invariant one-forms on the $S^3$.  In terms of the Euler angles $(\bar{\psi}, \bar{\phi},\bar{\theta})$
\be
\sigma_1= \cos \bar{\psi} \, d \bar{\theta} + \sin{\bar{\psi}} \, \sin{\bar{\theta}} \, d \bar{\phi}
\ee
\be
\sigma_2= -\sin{\bar{\psi}} \, d \bar{\theta} + \cos{\bar{\psi}} \,  \sin{\bar{\theta}} \, d \bar{\phi}
\ee
\be
\sigma_3= d \bar{\psi} + \cos{\bar{\theta}} \, d \bar{\phi}
\ee
and the Euler angles are given in terms of conventional parameters $(\psi, \phi, \theta)$ (e.g. Myers-Perry \cite{MP}) by
\be
\bar{\psi} = \psi + \phi, \, \, \, \, \,  \, \, \, \, \bar{\phi} = \phi - \psi,  \, \, \, \, \, \, \, \, \, \bar{\theta} = 2 \theta
\ee
The metric (\ref{ansatz}) in terms of the Euler angles is
\be \label{ansatz2}
ds^2 =  \frac{dr^2}{W(r)} + \alpha(r) \Big(d \bar{\theta}^2 + \sin^2{\bar{\theta}} \, d\bar{\phi}^2 \Big) + \beta(r) \Big({d \bar{\psi} + \cos{\bar{\theta}} \, d\bar{\phi}}\Big)^2
\ee
or in terms of the usual angles on the three sphere
$$
ds^2 = \frac{dr^2}{W(r)}+ 4 \alpha(r) d \theta^2 + 4 \sin^2{\theta} \, [\beta(r) \sin^2{\theta} + \alpha(r) \cos^2{\theta} ] d\psi^2
$$
$$
+ 4 \cos^2{\theta} \, [ \beta(r) \cos^2{\theta} + \alpha(r) \sin^2{\theta} ] d\phi^2
$$
\be \label{ansatz3}
 + 8 \sin^2{\theta} \cos^2{\theta} \, (\beta(r) - \alpha(r)) d\psi d\phi
\ee
In particular note that $\alpha = \beta = r^2/4$ gives an undistorted $S^3$.  On the other hand, if $\beta$ has a simple zero and $\alpha > 0$ in the region where $\beta \geq 0$ the metric (\ref{ansatz}) will describe a bubble of nothing.

In order to  contruct time symmetric initial data for a spacetime of dimension d with cosmological constant $\Lambda$ one need only require that the scalar curvature ${}^{(d-1)} R  = 2 \Lambda = -(d-1) (d-2)/l^2$.  While it will be convenient throughout this work to describe solutions in terms of a finite AdS length $l$, the corresponding asymptotically flat solutions may always be recovered by taking the limit $l \rightarrow \infty$.   For five dimensions with the present ansatz (\ref{ansatz2}) the constraint implies
\be \label{constraint}
\Big[ \frac{\alpha'}{\alpha} + \frac{\beta'}{2 \beta} \Big] W' + 2 \Big[\frac{\alpha''}{\alpha} + \frac{\beta''}{2 \beta} + \frac{\alpha' \beta'}{2 \alpha \beta} - \frac{{\alpha'}^2}{4 \alpha^2} - \frac{{\beta'}^2}{4 \beta^2}  \Big] W = \frac{12}{l^2} - \frac{\beta}{2\alpha^2} + \frac{2}{\alpha}
\ee
which is to say the constraint is simply a linear first order differential equation for $W(r)$.  This may be immediately solved for any $\alpha$ and $\beta$:
\be \label{sol1}
W = \frac{2 \beta e^{-\int_{r_0}^{r} ds J(s)}}{(r \beta' + 2 \beta \frac{r \alpha'}{\alpha})^2} \Big[C_1 + \int_{r_0}^{r} d\bar{r} \bar{r}^2 \Big(\frac{2}{\alpha} - \frac{\beta}{2 \alpha^2} + \frac{12}{l^2} \Big) \Big(\beta' + 2 \frac{\beta \alpha'}{\alpha} \Big)  e^{\int_{r_0}^{\bar{r}} dt J(t)} \Big]
\ee
where
\be
J(s) = \frac{\frac{3}{2} \Big(\frac{\alpha'}{\alpha} \Big)^2}{\frac{\alpha'}{\alpha} + \frac{\beta'}{2\beta}} - \frac{2}{s}
\ee
and $C_1$ is an integration constant which, as I will show shortly, will be fixed by requiring the solution to be regular.  Most of the interesting solutions to $(\ref{constraint})$ are bubble solutions containing a minimal $S^2$.    Specifically consider the case where $\beta$ has a simple zero at some radius $r_0$ while $\alpha(r_0)$ is a nonzero constant:
\be
\beta = k_0 (r - r_0) + \mathcal{O}\Big((r-r_0)^2\Big),  \, \, \, \, \, \, \alpha = k_1 + \mathcal{O} (r - r_0)
\ee
Then (\ref{sol1}) implies that
\be
W = \frac{2 C_1}{k_0 r_0^2} (r - r_0) + \mathcal{O}(r - r_0)^2
\ee
There are two components of the metric which look potentially troublesome in this case:
\be \label{nearbubblemet}
ds^2 = \frac{dr^2}{W(r)} + \beta(r) d\bar{\psi}^2 + \ldots
\ee
While the $g_{rr}$ component is diverging, this is purely a coordinate effect.  If one takes $x = \sqrt{r_0 (r - r_0)}$ (\ref{nearbubblemet}) becomes
\be
ds^2 = \frac{2 r_0 k_0}{C_1} (dx^2 + \frac{C_1}{2 r_0^2} x^2 d\bar{\psi}^2) + \ldots
\ee
Then the metric is regular near $r_0$ with the possible exception of a conical singularity.  Recalling that the period of $\bar{\psi}$ is $4 \pi$, this singularity will be eliminated if one takes 
\be
C_1 = \frac{r_0^2}{2}
\ee
More generically one could allow an orbifold singularity in the interior of the spacetime, in  which case one would have
\be
C_1 = \frac{r_0^2}{2 {n_1}^2}
\ee
for some integer $n_1$.  Taking $n_1 > 1$ makes the bubble heavier (in accordance with the experience that deficit angles are related to positive energies \cite{HawkingHorowitz}) but the remaining analysis will largely go through as for $n_1 = 1$.  Hence for the remainder of this work I will restrict myself to $n_1 = 1$.
Note generically for (\ref{sol1}) to be valid one must require
\be \label{condit1}
r \beta' + 2 \beta \frac{r \alpha'}{\alpha} > 0
\ee
which is to say that if the coefficient of $W'(r)$ in (\ref{constraint}) vanishes the corresponding solution to the constraint will typically be singular.  It will turn out this condition physically ensures the absence of any apparent horizons.   One must further require that that $W(r)$ never becomes negative.  A convenient sufficient, but not necessary, condition is to require (for non-positive cosmological constant)
\be \label{condit2}
\frac{2}{\alpha} - \frac{\beta}{2 \alpha^2} + \frac{12}{l^2} > 0
\ee
for then the constraint (\ref{constraint}) along with (\ref{condit1}) will ensure that if $W(r_1) = 0 $, then $W'(r_1) > 0$.  A violation of (\ref{condit2}), on the other hand, implies not that the solution is singular but rather that the case must be investigated on an individual basis.

While the solution has been written in terms of the AdS length scale $l$  solutions in deSitter space can also be obtained by taking $l^2 \rightarrow -l^2$, but I will leave the discussion of these solutions for future work.  Finally note that one could get topologically trivial regular solutions by requiring $\alpha, \beta \rightarrow \frac{r^2}{4}$ near the $r=0$ and formally taking the limit $r_0 \rightarrow 0$ in (\ref{sol1}).  While generically these solutions are merely perturbations of asymptotically flat and AdS space, in the AdS case it will turn out a subset of these perturbations will correspond to some interesting behavior in the boundary stress tensor.

Now turning to the asymptotics of the solutions, for the asymptotically flat case one should take
\be
\alpha(r) = \frac{r^2}{4} \Big[1 - \frac{a_0}{r^2} + \mathcal{O}\Big(\frac{1}{r^{2 + \epsilon}} \Big) \Big]
\ee
and
\be
\beta(r) = \frac{r^2}{4} \Big[1 - \frac{a_1}{r^2} + \mathcal{O}\Big(\frac{1}{r^{2 + \epsilon}} \Big) \Big]
\ee
for some constants $a_0$ and $a_1$.   Then the constraint implies
\be
W(r) = 1 - \frac{a_2}{r^2} +  \mathcal{O}\Big(\frac{1}{r^{2 + \epsilon}}, \frac{1}{r^4} \Big)
\ee
for some constant $a_2$, which is to say one has the usual asymptotically flat conditions.   For the asymptotically AdS case, provided one takes
\be
\alpha(r) = \frac{r^2}{4} \Big[1 - \frac{b_0}{r^4} + \mathcal{O}\Big(\frac{1}{r^{4 + \epsilon}} \Big) \Big]
\ee
and
\be
\beta(r) = \frac{r^2}{4} \Big[1 - \frac{b_1}{r^4} + \mathcal{O}\Big(\frac{1}{r^{4 + \epsilon}} \Big) \Big]
\ee
for some constants $b_0$ and $b_1$
the solution implies
\be
W(r) = \frac{r^2}{l^2} + 1 - \frac{b_2}{r^4} + \mathcal{O}\Big(\frac{1}{r^{4 + \epsilon}}, \frac{1}{r^6} \Big)
\ee
for some constant $b_2$ which means the solutions are asymptotically AdS in, for example, the sense of Henneaux and Teitelboim \cite{HenneauxTeitelboim}.

Clearly the original ansatz (\ref{ansatz}) does not entirely fix the gauge freedom.   One could have taken $W(r) = 1$ but then the constraint (\ref{constraint}) would be a nonlinear equation without any apparent explicit solution.   Provided that $\alpha$ is monotonically increasing, one looses no generality in taking
\be
\alpha(r) = \frac{r^2}{4}
\ee
Note this $r$ then smoothly interpolates between the radius of an $S^3$ and that of an $S^2$ (or to be precise near the bubble is proportional to that of an $S^2$).   Unless otherwise specified, the remainder of this work will be carried out in this gauge.  It is further convenient to factor out the long distance behavior of both $\beta$ and $W(r)$ by defining
\be
\beta(r) = \frac{r^2}{4} \gamma(r)
\ee
and
\be
W(r) = \Big(\frac{r^2}{l^2} + 1 - \frac{U(r)}{r^2} \Big) \gamma(r)
\ee
Note then one can describe each bubble solution, via $\gamma$, as a particular way of interpolating between zero and one.  With this gauge fixing and change of variables 
$$
U = \frac{4 r_0^2 e^{\int_{r_0}^{r} ds \frac{2 \gamma'}{6 \gamma +  s \gamma'}}}{(6 \gamma + r \gamma')^2} \Bigg[ \Big(\frac{r_0 \gamma'(r_0)}{2} \Big)^2 \Big(\frac{r_0^2}{l^2} + 1 \Big) - 1
$$
$$
+ \int_{r_0}^{r} \frac{d\bar{r} \, \bar{r}}{2 r_0^2} (6 \gamma + \bar{r} \gamma')\Big[ (\gamma - 1)\Big(8 + 12 \frac{r^2}{l^2} \Big) + \bar{r} \gamma' \Big(7 + 8 \frac{r^2}{l^2} \Big)
$$
\be \label{sol2c}
  + \bar{r}^2 \gamma'' \Big(1 +  \frac{r^2}{l^2} \Big) \Big] e^{-\int_{r_0}^{\bar{r}} dt \frac{2 \gamma'}{6 \gamma +  t \gamma'}} \Bigg]
\ee
The condition (\ref{condit1}) which, in addition to the requirement that $\gamma > 0$ for $r > r_0$, generically insures $U$ is a valid solution becomes
\be \label{condit3}
6 \gamma + r \gamma' > 0
\ee
In fact (\ref{condit3}) will assure that $\gamma > 0$ provided $\gamma'(r_0) > 0$; if $\gamma$ went through a second zero $r_1$ then $\gamma'(r_1) < 0$ and (\ref{condit3}) would be violated.   The previous criterion (\ref{condit2}) which insures $W(r) > 0$ for $r > r_0$ becomes
\be \label{condit4}
\gamma < 4 + \frac{6 r^2}{l^2}
\ee
which is to simply say that $\gamma$ should not become too large.

 \setcounter{equation}{0}
 \section{Geometric Properties and Dynamics}

 \subsection{Initial Dynamics}
 
 While the full dynamical evolution of this initial data apparently requires numerical attention, analytically one may at least discuss the initial time behavior.   For time symmetric initial data at the moment of time symmetry($t = 0$) using the Hamiltonian evolution equations with vanishing shift ($N^{a} = 0$) one finds
\be \label{timedeph2}
\ddot{h}_{a b}(0) =  -2 \, N^2 \Big[\, \,  {}^{(d - 1)}R_{a b} +  \frac{d-1}{l^2} \, h_{a b} (0) \Big] + 2 N D_a D_b N 
\ee
where $h_{a b}$ is the spatial metric, $D_a$ the derivative compatible with it, and all quantities are evaluated at the moment of time symmetry.
Note also any zeroes of an angle, such as $\bar{\psi}$, remain at the same coordinates (at least through second order in t) since
\be
\ddot{h}_{\bar{\psi} \bar{\psi}} = \ddot{h}_{a b} \Big(\frac{\partial}{\partial \bar{\psi}}\Big)^{a} \Big (\frac{\partial}{\partial \bar{\psi}}\Big)^{b}   = 0
\ee
for any regular initial data since then all the quantities on the right hand side of (\ref{timedeph2}) are regular.  Then since the area of the bubble is given, at least through order $t^2$, by
\be
A = \int d\bar{\theta} d\bar{\phi} \sqrt{h_{\bar{\theta} \bar{\theta}} h_{\bar{\phi} \bar{\phi}}}
\ee
evaluated at the bubble radius $r = r_0$ the acceleration of the area is
\be
\ddot{A} = \int d\bar{\theta} d\bar{\phi} \frac{\ddot{h}_{\bar{\theta} \bar{\theta}} h_{\bar{\phi} \bar{\phi}} + h_{\bar{\theta} \bar{\theta}} \ddot{h}_{\bar{\phi} \bar{\phi}} }{2 \sqrt{h_{\bar{\theta} \bar{\theta}} h_{\bar{\phi} \bar{\phi}}}}
\ee
If one takes $N$ to respect the $S^1 \times S^2$ symmetry (i.e. $N(t,r)$) then the terms involving derivatives of $N$ vanish and hence
\be
\ddot{A} = -N^2 \int d\bar{\theta} d\bar{\phi} \frac{ {{}^{(d-1)} R}_{\bar{\theta} \bar{\theta}} h_{\bar{\phi} \bar{\phi}} + h_{\bar{\theta} \bar{\theta}} \, \, {{}^{(d-1)} R}_{\bar{\phi} \bar{\phi}} }{\sqrt{h_{\bar{\theta} \bar{\theta}} h_{\bar{\phi} \bar{\psi}}}} - \frac{2 (d - 1)}{l^2} N^2 A
\ee
Then, for the particular bubbles being considered (\ref{ansatz})
\be \label{initexp}
\ddot{A} = 8 \pi N^2 \Big( \frac{1}{r_0 \gamma'(r_0)} - 1 - \frac{r_0^2}{l^2} \Big)
\ee
Hence one may find bubbles which begin expanding at any desired rate by making $r_0 \gamma'(r_0)$ appropriately small.

More generically, since evolution can not break the $S^1 \times S^2$ symmetry, the time dependent metric may be written
\be \label{timedepmet}
ds^2 =  g_{tt} (r,t) dt^2 + \frac{dr^2}{W(r,t)} + \alpha(r,t) \Big(d \bar{\theta}^2 + \sin^2(\bar{\theta}) d\bar{\phi}^2 \Big) + \beta(r,t) \Big({d \bar{\psi} + \cos(\bar{\theta}) d\bar{\phi}}\Big)^2
\ee
Given (\ref{timedepmet}) one can approximate the evolution of the spacetime in some neighborhood of the initial data surface by writing a perturbation series in powers of $t$.   In particular one may find the time dependence of the metric at the surface of a bubble beyond the leading order term described above.  This will prove useful for some examples described later.

One may also expand the time dependent spacetime around the bubble via (\ref{timedepmet}).   In particular, if one takes a gauge where
\be
 g_{t t} (r,t) = g_{t t} (r) = t_0 + t_1 (r - r_0) + \mathcal{O}\Big((r-r_0)^2 \Big)
 \ee
and defines $u(t) = \sqrt{\alpha(r_0,t)}$ and $v(t) = \sqrt{\beta'(r_0,t)}$ where a prime denotes a derivative with respect to r, one finds
\be \label{nearbubbleexp}
u''(t) = -u(t) \Big[ \frac{v''(t)}{v(t)} + \frac{t_1}{4 v^2(t)} - 2\frac{t_0}{l} \Big]
\ee
Hence generically a rapidly expanding bubble leads to a rapidly contracting $S^1$ and vice versa.  Furthermore, (\ref{nearbubbleexp}) suggests several interesting special cases.   If $v(t)$ is a constant then $u(t) = k_0 \cosh (c_0 t)$ for some constants $k_0$ and $c_0$.  Note $c_0$ will be real or imaginary according to whether the bubble begins expanding or collapsing (\ref{initexp}).   In the former case the bubble necessarily expands forever, or at least until the evolution breaks down (e.g. by the bubble hitting null infinity or one encounters a naked singularity).   It is also interesting to note in this case the metric on the bubble is equivalent to Witten's bubble.  More generically, if, in a gauge where $t_1 = 0$, $v(t)$ is proportional to $\cos(c_1 t)$ the solution for $u(t)$ will be of the same form .  In the later case, however, if $c_1$ is real it is not inconceivable that $v(t)$ goes through a zero before any interruption in the evolution and presumably produces a naked singularity.   More generically one could switch the roles of $r$ and $t$ in the evolution problem and specifying one component of the metric on the bubble (namely $u(t)$ or $v(t)$) arbitrarily will determine the spacetime.  In particular then one could specify the time dependence of the bubble to be anything one liked.  Analytically, however, it is difficult to tell whether any particular choice of $u(t)$ or $v(t)$, including the special cases mentioned above, correspond to regular initial data with physically interesting asymptotics.  It would be interesting to investigate this issue numerically.

\subsection{Static Bubbles}

It is natural to ask whether any of the described bubbles are static.  In fact there are static regular bubbles, but, as described below in detail, only with modified asymptotics.  The static solutions may be written as
\be
ds^2 =  -e^{-\int_{r}^{\infty} ds \lambda(s)}  dt^2 + \frac{dr^2}{W(r)} + \frac{r^2}{4} \Big(d \bar{\theta}^2 + \sin^2(\bar{\theta}) d\bar{\phi}^2 \Big) + \frac{r^2}{4} \gamma (r)  \Big({d \bar{\psi} + \cos(\bar{\theta}) d\bar{\phi}}\Big)^2
\ee
where the Einstein equations become
\be \label{statgamma}
\gamma'' = \frac{{\gamma'}^2}{\gamma} - \frac{\gamma'}{r} + \frac{8 \gamma^2 - 4 \Big(1 + \frac{r^2}{l^2} \Big) r \gamma' + 2 \gamma (r \gamma' - 4)}{r^2W(r)}
\ee
$$
W' = \frac{1}{r (6\gamma + r \gamma')}  \Big[-20 \gamma^2 + 8 \Big(1+\frac{r^2}{l^2} \Big) r \gamma' + 4  \Big(8 + 6 \frac{r^2}{l^2} - r \gamma' \Big) \gamma
$$
\be \label{statW}
- \Big(12 \gamma + 6 r \gamma' + \frac{r^2 {\gamma'}^2}{\gamma}\Big) W \Big]
\ee
and
\be \label{Statg}
\lambda =  \frac{4}{r (6\gamma + r \gamma')}  \Bigg[ - 3 \gamma - 4 r \gamma' + \frac{\gamma \Big(6 \frac{r^2}{l^2} + 4 - \gamma \Big)}{W} \Bigg]
\ee
Then (\ref{statgamma}) can be solved for $W(r)$ and (\ref{statW}) becomes a third order ordinary differential equation for $\gamma(r)$ which can be solved numerically.  Given $\gamma(r)$ one can then use  (\ref{statgamma})  and  (\ref{Statg}) to find $W(r)$ and $\lambda(r)$.  The precise equation for $\gamma(r)$ is, however, sufficiently lengthy and unilluminating that I will omit it here. 

If one demands that the bubble be regular, including that it lacks any conical singularity, two of the three parameters in solving the third order ODE for $\gamma$ are eliminated.  One may take the remaining parameter as the size of the bubble $r_0$.  For the asymptotically flat case, of course, this family just corresponds to constant conformal rescalings of the metric but in the AdS case various properties of the metric depend on the ratio $r_0/l$.

Consider first the asymptotically flat case.  One may numerically solve for the described functions, but in fact one such bubble may be constructed analytically by lifting the (Euclidean) Taub-Bolt solution \cite{Page1}.   Since the above equations admit only one regular bubble, numerically one simply finds the lifted Taub-Bolt, as one can verify in detail.

Now consider the AdS case.  For the usual AdS asymptotics, $\gamma \rightarrow 1$  at large r .   However, for these static regular bubbles instead one finds that asymptotically $\gamma \rightarrow c$ where $c$ is a relatively small number.  In fact,  the maximum $\gamma(\infty) \approx 0.0765$ occurs at $r_0 \approx 0.85 l $.  See Figure 1.   Numerical investigation shows the remaining metric functions approach their standard values, at least at leading order; at large r
\be
W(r) \approx \frac{r^2}{l^2}
\ee
and
\be
g_{t t} (r) = -e^{-\int_{r}^{\infty} ds \lambda(s)}  \approx -\frac{r^2}{l^2}
\ee
\begin{figure}
\begin{picture} (0,0)
    	\put(-125,5){$\gamma(\infty)$}
         \put(142, -160){$\frac{r}{l}$}
    \end{picture}
    \centering

	\includegraphics[scale= 1]{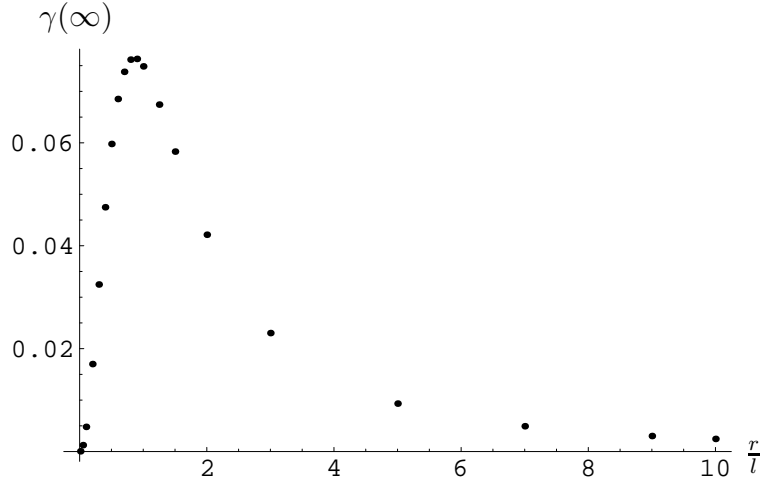}
	\caption{Asymptotic value of $\gamma$ for various size static AdS bubbles}
	\label{AdSStatasym}
	\end{figure}
	
\begin{figure}
\begin{picture} (0,0)
    	\put(22,2){$\gamma$}
         \put(188, -106){$\frac{r}{l}$}
         
         \put(216,1){$\gamma$}
         \put(387, -114){$\frac{r}{l}$}

    \end{picture}

    \begin{minipage} [b] {.5\linewidth}
	\includegraphics[scale= .8]{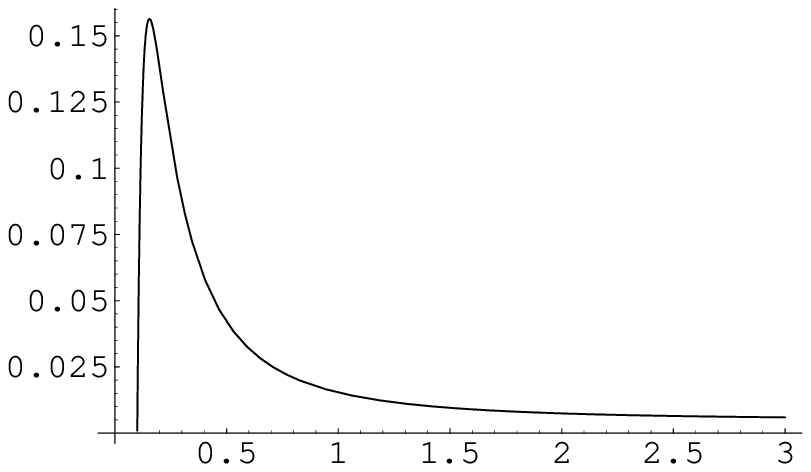}
	\caption{$\gamma(r)$ for $r_0 = \frac{l}{10}$}

    \label{gammasmall}

\end{minipage} %
    \begin{minipage} [b] {.5\linewidth}
	\includegraphics[scale=.8]{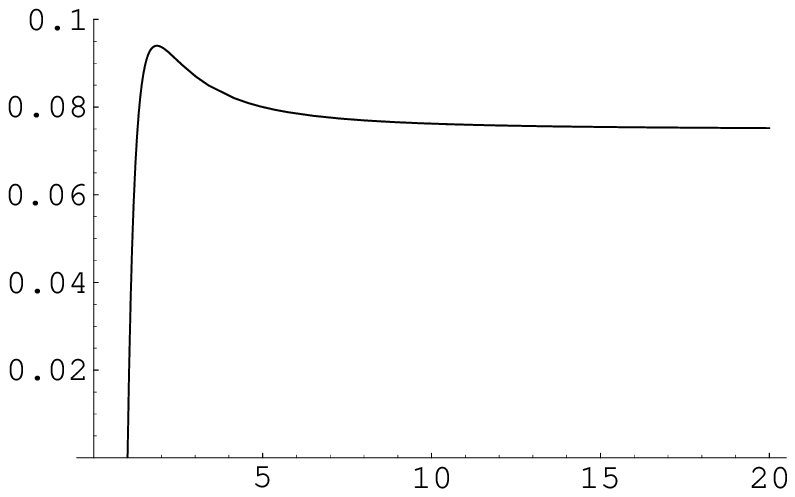}
	\caption{$\gamma(r)$ for $r_0 = l$}

    \end{minipage}
\end{figure}

	This is to say the asymptotic sphere, while topologically a $S^3$, is squashed instead of round.   In this case the asymptotically AdS space qualitatively bears some resemblance to AdS with a Kaluza-Klein direction, in which case solitons have been previously discussed \cite{MHC}.   The fact, however, that there is a purely gravitational solution with a boundary topologically equivalent to global AdS is, however, somewhat unexpected.  The solution may also qualitatively be viewed as a ``lift'' to five dimensions of the Euclidean AdS-Taub-Bolt solution \cite{Page2}.  While this characterization is necessarily somewhat rough, it is of some interest to note that the Euclidan AdS-Taub-Bolt solutions involve highly squashed spheres as well; rewriting the globally AdS Taub-Bolt solutions in terms of the spatial metric as above the maximum value of $\gamma(\infty)$ is
	\be
	\frac{1}{3 (2 + \sqrt{3})} \approx 0.0893
	\ee
See Figures \ref{gammasmall}-\ref{gttlarge} for the metric functions for small, medium, and large AdS static bubbles.  In particular note that, perhaps contrary to one's expectations, $\gamma$ rises rather rapidly near the bubble and is often not even monotonic.

\begin{figure}
\begin{picture} (0,0)
     	\put(-65,2){$\gamma$}
         \put(96, -107){$\frac{r}{l}$}
         
        \end{picture}
    \centering

	\includegraphics[scale= .8]{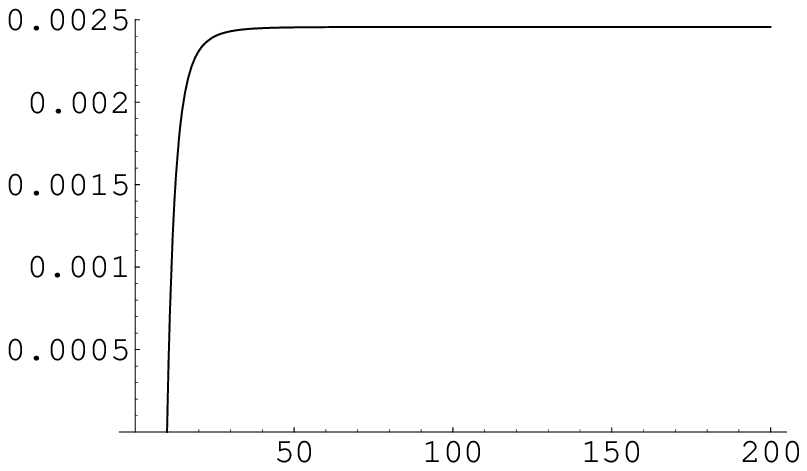}
	\caption{$\gamma(r)$ for $r_0 = 10 l$}
	
	\end{figure}
	
	\begin{figure}
\begin{picture} (0,0)
    	\put(5,3){$\frac{W}{\frac{r^2}{l^2} + 1}$}
         \put(192, -120){$\frac{r}{l}$}
         
         \put(202,10){$\frac{W}{\frac{r^2}{l^2} + 1}$}
         \put(385, -110){$\frac{r}{l}$}

    \end{picture}

    \begin{minipage} [b] {.5\linewidth}
	\includegraphics[scale= .8]{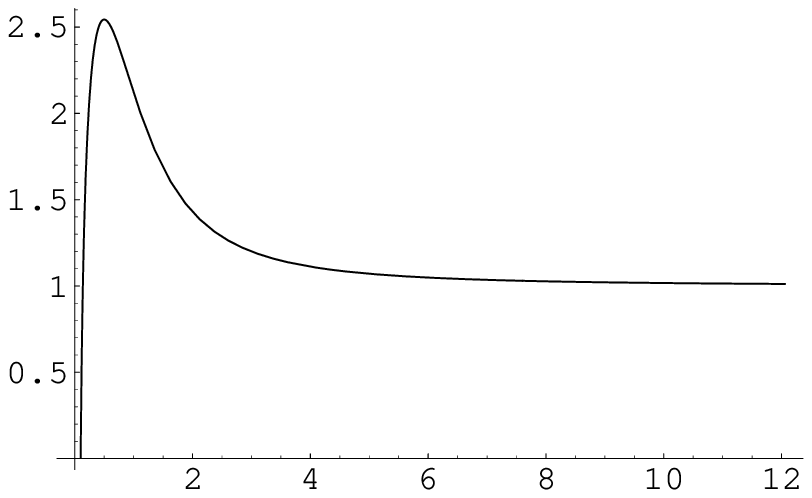}
	\caption{$W(r)$ for $r_0 = \frac{l}{10}$}

\end{minipage} %
    \begin{minipage} [b] {.5\linewidth}
	\includegraphics[scale=.8]{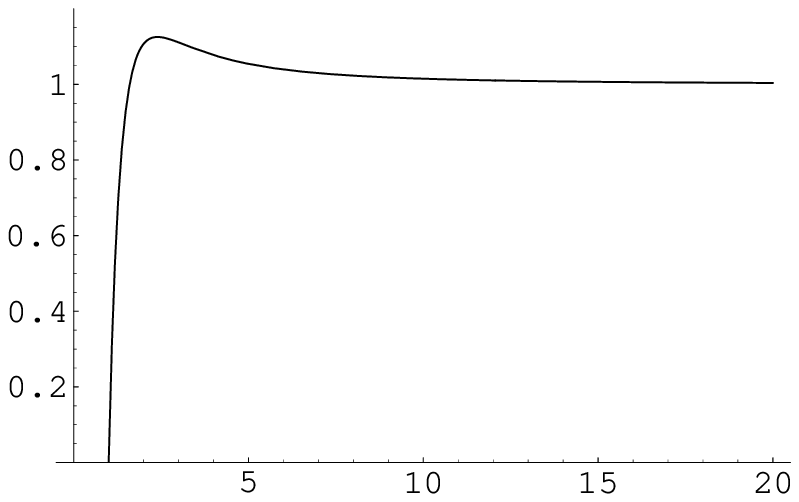}
	\caption{$\frac{W(r)}{\frac{r^2}{l^2} + 1}$ for $r_0 = l$}
    \label{figexp9}

    \end{minipage}
\end{figure}

\begin{figure}
\begin{picture} (0,0)
     	\put(-88,10){$\frac{W}{\frac{r^2}{l^2} + 1}$}
         \put(95, -110){$\frac{r}{l}$}
         
        \end{picture}
    \centering

	\includegraphics[scale= .8]{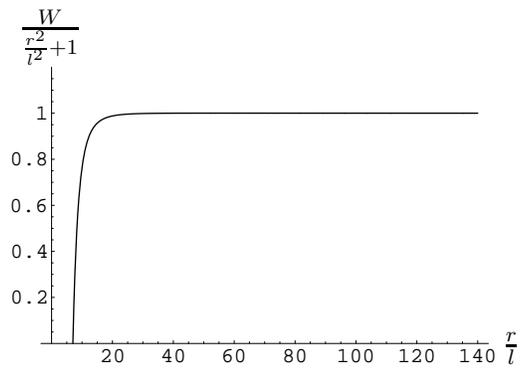}
	\caption{$\frac{W(r)}{\frac{r^2}{l^2} + 1}$ for $r_0 = 10 l$}
	
	\end{figure}
	
	\begin{figure}
\begin{picture} (0,0)
    	\put(13,2){$\lambda$}
         \put(190, -108){$\frac{r}{l}$}
         
         \put(210,2){$\lambda$}
         \put(383, -108){$\frac{r}{l}$}

    \end{picture}

    \begin{minipage} [b] {.5\linewidth}
	\includegraphics[scale= .8]{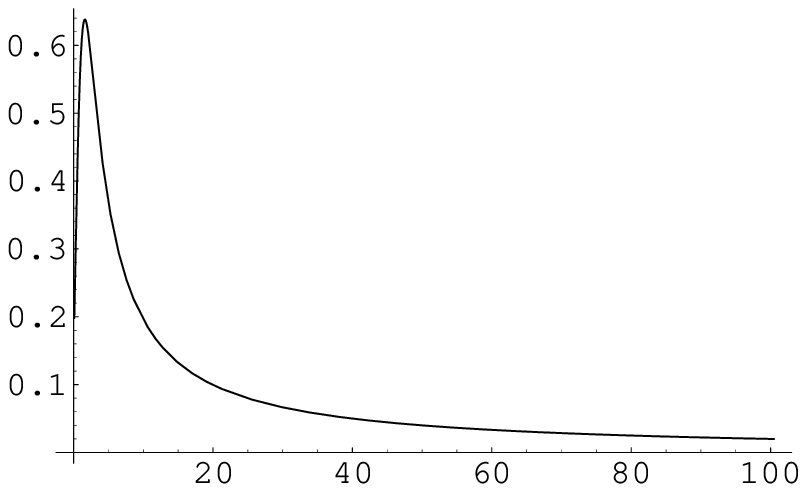}
	\caption{$\lambda (r)$ for $r_0 = \frac{l}{10}$}

\end{minipage} %
    \begin{minipage} [b] {.5\linewidth}
	\includegraphics[scale=.8]{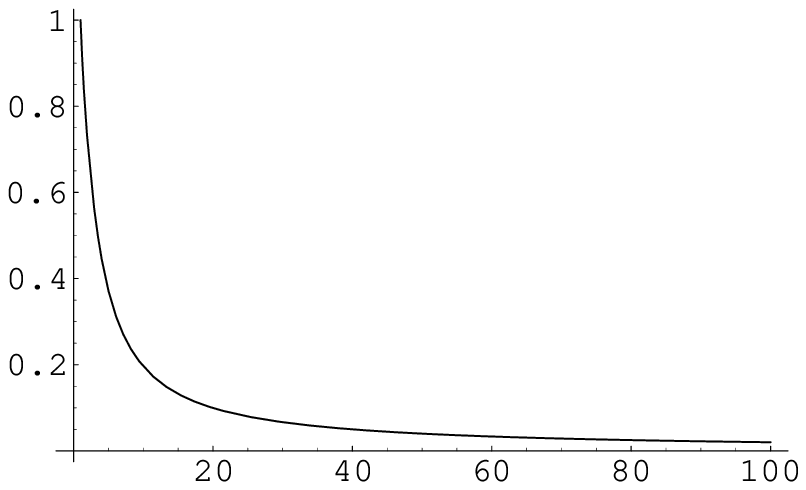}
	\caption{$\lambda (r) $ for $r_0 = l$}

    \end{minipage}
\end{figure}

\begin{figure}
\begin{picture} (0,0)
     	\put(-75,2){$\lambda$}
         \put(95, -107){$\frac{r}{l}$}
         
        \end{picture}
    \centering

	\includegraphics[scale= .8]{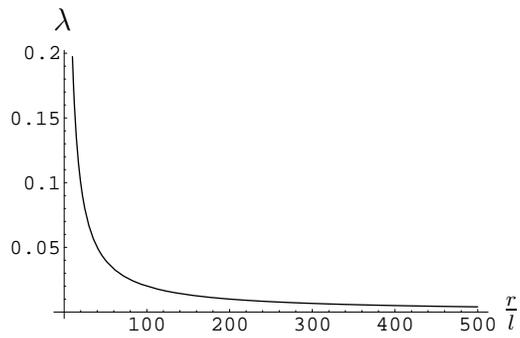}
	\caption{$\lambda(r)$ for $r_0 = 10 l$}
	\label{gttlarge}
	
	\end{figure}

	One can also find topologically trivial static solutions to (\ref{statgamma})-(\ref{Statg}).   Besides the spherically symmetric trivial case, all have distorted asymptotics qualitatively similar to that described above.  For the asymptotically flat case one has, as expected, the lift of the (self-dual) Taub-Nut solution \cite{Hawking}.   For the AdS case the question of whether in a given asymptotic class (whatever that might mean precisely) there are both static bubbles and topologically trivial solutions will be left for future work.

 \subsection{Apparent Horizons}
 
An apparent horizon is a marginally outer trapped surface where the expansion $\theta$ of the outgoing family of null geodesics vanishes, i.e.
\be
\theta = \nabla_a l^a = 0
\ee
where $l$ is an outgoing null geodesic.   In terms of initial data on a surface $\Sigma$ one finds
\be
\theta = D_a s^a + K - s^a s^b K_{a b}
\ee
where $D_a$ is the covariant derivative compatible with the induced metric on $\Sigma$, $K_{a b}$ is the extrinsic curvature of $\Sigma$ and $s^a$ is a spacelike unit normal to the surface in question.  In particular, for time symmetric initial data an apparent horizon is present only if
\be
D_a s^a = 0
\ee
The bubbles under consideration have an $S^1 \times S^2$ symmetry so any apparent horizon must be at constant r.   Then plugging in the initial data (\ref{ansatz2}) one finds
\be
D_a s^a = \sqrt{W} \Big(\frac{\alpha'}{\alpha} + \frac{\beta'}{2 \beta} \Big)
\ee
and so is positive definite provided (\ref{condit1}) is imposed.   On the other hand if (\ref{condit1})  were to be violated one would have an apparent horizon at the outermost such radius.  For the sake of comparison, note that if one applied these formulas to the standard stationary black holes one would have an apparent horizon at the outermost zero of $W$.  This is, of course, precisely as one expects.

 \subsection{Energy}
 
Given a spacelike slice $\Sigma$ in an asymptotically AdS space, the energy may be given with an ADM-type definition as the value of the Hamiltonian which becomes
 \be \label{engeneral}
E = \frac{1}{16 \pi G} \int dS^{a} \Big[ N D^b (\delta h_{a b}) - D^b(N) \delta h_{a b} + h^{c d}(-N D_a (\delta h_{c d}) + D_a (N) \delta h_{c d} \Big]
\ee
where the integral is over over the (asymptotic) spatial boundary of $\Sigma$, $h_{a b}$ is the spatial metric induced on $\Sigma$ and $\delta h_{a b} = h_{a b} - h^{0}_{a b}$ where $h^{0}_{a b}$ is the spatial metric induced by a background (in this case pure AdS) metric on $\Sigma$.  The second and the fourth terms may be somewhat unfamiliar and, indeed, in the asymptotically flat case they vanish.   However, for AdS the value of the lapse asymptotically grows ($N \sim r/L$) and hence these terms are significant.  This definition agrees with that of Henneaux and Teitelboim \cite{HenneauxTeitelboim} and hence with essentially all definitions of the energy in AdS \cite{HollandsIshibashiMarolf}. \footnote{The reader familiar with the details of \cite{HawkingHorowitz} might be confused by this statement, but in the case where one can imbed the metric isometrically one can choose a gauge where the terms involving the gradients of $N$ vanish and these authors have done so.}  If one further considers line elements of the type
$$
ds^2 = \frac{dr^2}{\frac{r^2}{l^2} + 1 - \delta_{r r} \frac{l^2}{r^{2}} + \mathcal{O}\Big(\frac{1}{r^{2 + \epsilon}} \Big)} + r^2 \Big(1 + \delta_{\theta \theta} \frac{l^4}{r^{4}}  + \mathcal{O}\Big(\frac{1}{r^{4 + \epsilon}} \Big) \Big) d\theta^2 
$$
$$+ r^2 \sin^2(\theta) \Big(1 + \delta_{\psi \psi} \frac{l^4}{r^{4}} + \mathcal{O}\Big(\frac{1}{r^{4 + \epsilon}} \Big) \Big) d\psi^2 +r^2 \cos^2(\theta) \Big(1 +\delta_{\phi \phi} \frac{l^4}{r^{4}} + \mathcal{O}\Big(\frac{1}{r^{4 + \epsilon}} \Big) \Big) d\phi^2$$
\be
 + 2 \delta_{\psi \phi} \frac{l^4}{r^2} \sin^2(\theta) \cos^2(\theta) d\psi d\phi
\ee
Then
\be \label{EAdsdef}
E_{AdS} = \frac{l^2}{16 \pi G} \int d\theta d\psi d\phi \sin(\theta) \cos(\theta) \Big[3 \delta_{rr} +4 (\delta_{\theta \theta} + \delta_{\psi \psi} + \delta_{\phi \phi} ) \Big]
\ee
Note for the bubble solutions in the gauge $\alpha = r^2/4 $ and
\be
\gamma = 1 + a_0 \frac{l^4}{r^4} +  \mathcal{O}\Big(\frac{1}{r^{4 + \epsilon}}  \Big)
\ee
then
\be
\delta_{\theta \theta} + \delta_{\psi \psi} + \delta_{\phi \phi} = a_0
\ee
while
\be
 \delta_{rr} = \frac{U(\infty)}{l^2} - a_0
 \ee

For the asymptotically flat case for line elements of the type
$$
ds^2 = \Big( 1 + \delta_{r r} \frac{r_0^2}{r^{2}} + \mathcal{O}\Big(\frac{1}{r^{2 + \epsilon}} \Big) \Big) dr^2+ r^2 \Big(1 + \delta_{\theta \theta} \frac{r_0^2}{r^{2}}  + \mathcal{O}\Big(\frac{1}{r^{2 + \epsilon}} \Big) \Big) d\theta^2 
$$
$$+ r^2 \sin^2(\theta) \Big(1 + \delta_{\psi \psi} \frac{r_0^2}{r^{2}} + \mathcal{O}\Big(\frac{1}{r^{2 + \epsilon}} \Big) \Big) d\psi^2 +r^2 \cos^2(\theta) \Big(1 + \delta_{\phi \phi} \frac{r_0^2}{r^{2}} + \mathcal{O}\Big(\frac{1}{r^{2 + \epsilon}} \Big) \Big) d\phi^2$$
\be
 + 2 r_0^2  \delta_{\psi \phi} \sin^2(\theta) \cos^2(\theta) d\psi d\phi
\ee
where $r_0$ is some convenient length scale (e.g. the bubble size) the ADM mass is simply
\be
E_{ADM} =   \frac{r_0 ^2}{16 \pi G} \int d\theta d\psi d\phi \sin(\theta) \cos(\theta) \Big[3 \delta_{rr} +  (\delta_{\theta \theta} + \delta_{\psi \psi} + \delta_{\phi \phi} ) \Big]
\ee
For bubbles in the gauge $\alpha = r^2/4$ if
\be
\gamma = 1 + b_0 \frac{r_0^2}{r^2} +  \mathcal{O}\Big(\frac{1}{r^{2+ \epsilon}} \Big)
\ee
then
\be
\delta_{\theta \theta} + \delta_{\psi \psi} + \delta_{\phi \phi}  = b_0
\ee
while $\delta_{rr} = \frac{U(\infty)}{r_0^2} - b_0$.

One would like to write $ U(\infty)$ in as illuminating a form as possible.  Performing several integrations by parts (\ref{sol2c}) becomes
$$
U(\infty) = \frac{r_0^2}{9} \Bigg[\lim_{r \rightarrow \infty} \frac{3 r^3 \gamma'}{r_0^2} \Big(1 + \frac{r^2}{l^2} \Big) + \lim_{r \rightarrow \infty} \frac{3 r^2}{r_0^2} (\gamma - 1) \Big(4 + 3 \frac{r^2}{l^2} \Big)
$$
$$
 + 3 \Big(1 +\frac{r_0^2}{2 l^2} \Big) e^{\int_{r_0}^{\infty} ds \frac{2\gamma'}{6 \gamma + s \gamma'}} + \int_{r_0}^{\infty} \frac{dr r}{r_0^2} e^{\int_{r}^{\infty} ds \frac{2\gamma'}{6 \gamma + s \gamma'}} \Big[ (\gamma-1)^2 \Big(8 + 6 \frac{r^2}{l^2} \Big)
 $$
 \be
  + \frac{r \gamma'}{6 \gamma + r \gamma'} \Big( 3 r \gamma \gamma' \Big(1 + \frac{r^2}{l^2} \Big) + (\gamma - 1) \Big(24 + 18 \frac{r^2}{l^2} \Big) + (\gamma - 1)^2 (16 + \frac{15 r^2}{l^2} )\Big) \Big] \Bigg]
\ee
As it turns out, the explicit asymptotic surface terms, which give the leading order behavior of $1 - \gamma$, cancel with the other explicit surface terms in the definition of the energy.   In the asymptotically flat case the energy becomes
$$
E_{ADM}  = \frac{2 \pi^2 r_0^2}{16 \pi G} \Bigg[ e^{\int_{r_0}^{\infty} ds \frac{2\gamma'}{6 \gamma + s \gamma'}} 
 $$
 \be \label{Eadmpos}
 + \int_{r_0}^{\infty} \frac{dr r}{r_0^2} e^{\int_{r}^{\infty} ds \frac{2\gamma'}{6 \gamma + s \gamma'}} \frac{ \gamma \Big(r \gamma' + 4 (\gamma - 1) \Big)^2}{6 \gamma + r\gamma'}\Bigg]
\ee
Note then this energy is positive definite, as expected from the positive energy theorems  \cite{WSY}.  Further, it is reasonably clear for the lightest bubbles
\be
E_{ADM} \sim  \frac{2 \pi^2}{16 \pi G} r_0^2
\ee
The only way one could make such bubbles very light would involve making the exponential in (\ref{Eadmpos}) very negative and by examining several examples one can convince oneself it is not possible to do this without producing other contributions which more than compensate for the small exponential.  Intuitively, this is as one naively expects; the given bubble has a $S^2$ symmetry and one would expect gradient energies on order set by the size of the bubble.  There are bubbles, however, which are substantially lighter than a black hole of the same size.  In particular, if one considers a $\gamma$ which rapidly goes from zero to one, then
\be \label{Eadmrapid}
E_{ADM} \approx \frac{2 \pi^2}{16 \pi G}  \frac{r_0^2}{2}
\ee
whereas for a black hole
\be
E_{BH} =  \frac{2 \pi^2}{16 \pi G} 3 r_+^2
\ee
where $r_+$ is the size of the horizon.   Then producing a black hole would require compressing all the energy of the spacetime into a region substantially smaller than the initial size of the bubble.
Since the expansion or contraction of the bubble is simply set by $\gamma'(r_0)$, one can definitely make expanding bubbles which are not simply black holes.  

	It should also be noted that (\ref{Eadmrapid}) is sufficiently light that even its collapse, immediate or eventual, is of some interest since one might expect an event horizon not to form until the bubble has collapsed in size an order of magnitude.  This suggests one might be able to start out with a bubble which is reasonably described classically but whose size might reach the string scale before a horizon is formed.   If such a phenomena did occur one would have a quantum version of cosmic censorship violation.  It would be interesting to investigate this possibility further.

For the AdS case the energy becomes
$$
E_{AdS} =  \frac{2 \pi^2 r_0^2}{16 \pi G} \Bigg[\Big(1 +\frac{r_0^2}{2 l^2} \Big) e^{\int_{r_0}^{\infty}  ds \frac{2\gamma'}{6 \gamma + s \gamma'}} 
$$
$$
+ \int_{r_0}^{\infty} \frac{dr r}{3 r_0^2r_0^2r_0^2} e^{\int_{r}^{\infty} \frac{2\gamma'}{6 \gamma + s \gamma'}} \Big[ (\gamma-1)^2 \Big(8 + 6 \frac{r^2}{l^2} \Big)
 $$
 \be \label{EAdSpos}
  + \frac{r \gamma'}{6 \gamma + r \gamma'} \Big(3 r \gamma \gamma' \Big(1 + \frac{r^2}{l^2} \Big) + (\gamma - 1) \Big(24 + 18 \frac{r^2}{l^2} \Big) + (\gamma - 1)^2 (16 + \frac{15 r^2}{l^2} ) \Big) \Big] \Bigg]
\ee
While (\ref{EAdSpos}) is not manifestly positive definite, the positive energy theorems assure one it will be \cite{PosEnergyAdS}.    One would like to know, however, whether such bubbles can be very light.   Given the above comments in regard to the asymptotically flat case, this is essentially equivalent to asking whether the integrand of the second term of (\ref{EAdSpos}) can ever become negative.  One sufficient, but not necessary, criterion is to consider whether for a given value of $\gamma$ whether there is any value of $r \gamma'$ such that the integrand ever becomes negative.  A bit of straightforward algebra shows that if $\gamma < 7 + 4 \sqrt{3} \approx 13.9$ the integrand is positive definite.    It is, however, possible to find cases for which $\gamma \gg 1$ produces a negative contribution to $(\ref{EAdSpos})$ for some region; intuitively this simply reflects that making a large volume in AdS gives a negative energy contribution from the cosmological constant.   Examining several examples provides evidence that the gradient energies produced in getting to $\gamma \gg 1$ always costs substantially more energy than it saves, although there does not seem to be any obvious proof of this conjecture.  In the absence of any such exception, then for large bubbles
\be
E_{AdS} \sim \frac{2 \pi^2}{16 \pi G} \frac{{r_0}^4}{l^2}
\ee
while for small bubbles (i.e. those for which $\gamma$ is substantially different from 1 only in a region $r \lesssim l$)
\be
E_{AdS} \sim  \frac{2 \pi^2}{16 \pi G} r_0^2
\ee
Again if one considers $\gamma$ which rapidly goes from zero to one then
\be
E_{AdS} \approx \frac{2 \pi^2}{16 \pi G}  \frac{r_0^2}{2} \Big(1 + \frac{r_0^2}{l^2} \Big)
\ee
wheras for a black hole in $AdS_5$ of size $r_+$
\be
E_{BH} = \frac{2 \pi^2}{16 \pi G}  3 {r_+}^2  \Big(1 + \frac{{r_+}^2}{l^2} \Big)
\ee
and again one can make both expanding and contracting bubbles substantially lighter than a comparable size black hole.

 \setcounter{equation}{0}
\section{Examples}

I have constructed time symmetric initial data describing bubbles of arbitrary size for any $\gamma$ obeying (\ref{condit3}) and (\ref{condit4}):
\be \label{ansatz4}
ds^2 =  \frac{dr^2}{\gamma(r) \Big(\frac{r^2}{l^2} + 1 - \frac{U(r)}{r^2} \Big)} + \frac{r^2}{4} \Big(d \bar{\theta}^2 + \sin^2{\bar{\theta}} \, d\bar{\phi}^2 \Big) + \frac{r^2}{4}  \gamma(r) \Big({d \bar{\psi} + \cos{\bar{\theta}} \, d\bar{\phi}}\Big)^2
\ee

While this generality is interesting in itself, it carries with it two costs.  The first is that the metric functions can almost never be written explicitly.   In particular the form of $U(r)$ (\ref{sol2c}) involves a double integration and it is difficult to find any $\gamma$ such that once the first integral is performed the second can be done explicitly.   Of course, one can always do the relevant integrals numerically and obtain physical quantities, such as the energy, in this manner.  The second cost is that there are so many solutions it is difficult to decide what the ``typical'' properties and behavior of bubbles are or how such properties translate into the features of $\gamma$.  One can, however, begin by examining a variety of examples and this section describes several of the more interesting ones.

\subsection{Example I}

For asymptotically flat space, there is in fact one case in which one can obtain $U(r)$ explicitly.  Consider
\be \label{Ex1}
\gamma = 1 - a_0 \Big(\frac{r_0}{r}\Big)^2 - a_1 \Big(\frac{r_0}{r}\Big)^4
\ee
Then one obtains a bubble of size $r_0$ if $a_1 = 1 - a_0$.  Further
\be
a_0 = 2 - \frac{r_0 \gamma'(r_0)}{2}
\ee
and hence $a_0 < 2$.  Note one has an expanding bubble if $\frac{3}{2} < a_0 < 2$ and a collapsing bubble if $a_0 < \frac{3}{2}$.   One must further check that $6 \gamma + r \gamma' > 0$, but a bit of algebra shows this is automatic once one imposes $\gamma'(r_0) > 0$.   The particular powers on (\ref{Ex1}) are chosen so that
\be
8 (\gamma - 1) +  7 r \gamma'  + r^2 \gamma'' = 0
\ee
and hence (\ref{sol2c}) $U(r)$ involves only a single integral.  One finds
$$
U(r) = \frac{r_0^2 [(2 - a_0)^2 - 1]}{\Big(3 - 2 a_0 \Big(\frac{r_0}{r}\Big)^2 -  a_1 \Big(\frac{r_0}{r}\Big)^4\Big)^2} \Bigg[ \Bigg(\frac{a_1 \Big(\frac{r_0}{r}\Big)^2 + a_0 + b_0}{a_1 + a_0 + b_0} \Bigg)^{1 + \frac{a_0}{2 b_0}}
$$
\be \label{simpU}
\times \Bigg(\frac{a_1 \Big(\frac{r_0}{r}\Big)^2 + a_0 - b_0}{a_1 + a_0 - b_0} \Bigg)^{1 - \frac{a_0}{2 b_0}} \Bigg]
\ee
where $b_0 =  \sqrt{{a_0}^2 +3 \, a_1}$.     Note in the case $a_0 = 1$ (see the overall prefactor in (\ref{simpU}) ) $U(r)$ vanishes.  In fact, this reproduces the single asymptotically flat example found by LeBrun \cite{LeBrun}.   That bubble is, however, not very exciting; it is not particularly light
\be
E = \frac{2 \pi^2}{16 \pi G} 2 r_0^2
\ee
and immediately collapses.  More generically one has
$$
E = \frac{2 \pi^2}{16 \pi G} \Big(3 U(\infty) + 2 a_0 r_0^2 \Big)
$$
\be
=  \frac{2 \pi^2 r_0^2}{16 \pi G} \Bigg[ \frac{1}{3} \Big((2-a_0)^2 - 1 \Big) \Big(\frac{a_1}{a_0 + b_0} + 1\Big)^{-1 - \frac{a_0}{2 b_0}} \Big(\frac{a_1}{a_0 - b_0} + 1\Big)^{-1 + \frac{a_0}{2 b_0}} + 2 a_0 \Bigg]
\ee
See Figure 11.

\begin{figure}
\begin{picture} (0,0)
    	\put(-40,3){E}
         \put(120, -133){$a_0$}
    \end{picture}
    \centering

	\includegraphics[scale= 1]{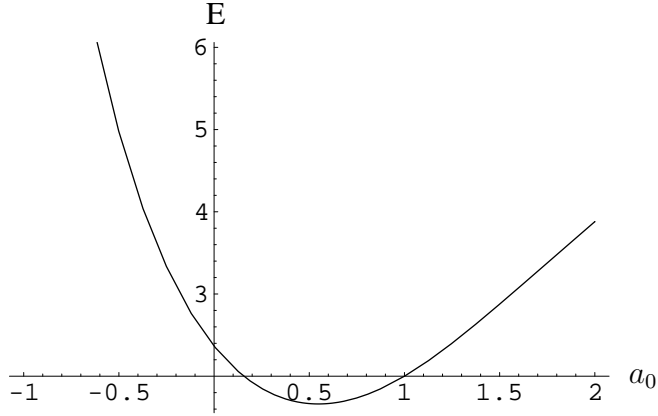}
	\label{Examp}
	\caption{Energy for example I}

	\end{figure}
Note that the minimum energy bubble within this family is initially collapsing and within this set of  expanding bubbles a larger initial expansion implies a larger energy.   There are two possible sets of intuitions relating expansion of bubbles and mass.  The first is that all of the bubbles described in this work are essentially gravitationally self bound and would naturally collapse.  However, one could, according to this point of view, create a tension in the spacetime to force these bubbles to expand but those gradients will cost energy and hence one would predict that the more rapidly expanding bubbles will be heavier.  Alternatively, one might suppose that there are bubbles which are not strongly bound by gravity but instead would prefer to relax away their gradient energies by expanding.   For such bubbles the expansion is held back by self-gravitation and hence one would expect that lighter bubbles will expand more rapidly.  The above example, as well as a variety of simple polynomial $\gamma$ in both the asymptotically flat and asymptotically AdS cases, show the former trend.  On the other hand, the asymptotically flat double bubbles \cite{CopseyBubbles1} and at least one set of Kaluza-Klein bubbles \cite{SarbachLehner} show the second trend.   It is in fact possible the described family of bubbles might be large enough to contain examples of both lines of argument.  Further, considering that the expressions for the mass (\ref{Eadmpos}), (\ref{EAdSpos}) make it difficult to determine the $\gamma$ which would minimize the mass for a given $\gamma'(r_0)$, there does not seem to be any simple way to determine whether the increase of mass of a given family of bubbles when a particular parameter is varied is primarily due to the changing initial acceleration of the bubble or the variation of the asymptotic metric (i.e. adding more gravitational radiation to the initial data).  Hence, it is unclear whether or not a polynomial example such as that discussed here is particularly natural or generic.

\subsection{Example II}

In the case of an AdS solution the only particularly simple interesting bubble appears to be
$$
ds^2 = -\Big(\frac{r^2}{l^2} + 1\Big) dt^2 + \frac{dr^2}{\Big(1 + \frac{r^2}{l^2}\Big)\Big(1 - \frac{r_0^4}{r^4}\Big)} + \frac{r^2}{4} \Big(d \bar{\theta}^2 + \sin^2(\bar{\theta}) d\bar{\phi}^2 \Big)
$$
\be \label{Ex2}
 +\frac{r^2}{4}\Big(1 - \frac{r_0^4}{r^4}\Big) \Big({d \bar{\psi} + \cos(\bar{\theta}) d\bar{\phi}}\Big)^2
\ee
While this metric is static, if asymptotically AdS or, taking the limit $l \rightarrow \infty$, asymptotically flat it suffers from a conical singularity; at the bubble the length of the $S^1$ is
\be
4 \pi \sqrt{1 + \frac{r_0^2}{l^2}} 
\ee
which is to say there is an excess angle.   If one were willing to quotient the $S^3$ one can find a smooth massless bubble in the asymptotically flat case and in the AdS case a regular bubble of mass
\be
M = -\frac{\pi^2}{16 \pi G} \frac{\frac{r_0^4}{l^2}}{\sqrt{1 + \frac{r_0^2}{l^2}}}
\ee
The quotiented asymptotically flat case does not appear to be of particular interest, especially since one knows from the work of LeBrun \cite{LeBrun} that there are negative mass solutions in that case.   However, AdS  with a quotient does provide an extension of  LeBrun's results to the case of negative cosmological constant.   In fact, these solutions have been found and studied previously and are conjectured to be the lowest energy state with such an AdS quotient \cite{ClarksonMann}.  Also note that a positive cosmological constant and a properly chosen $r_0$ provides a perfectly regular solution without any quotients.  

	One might wonder whether quantum effects might resolve this singularity.  For the AdS case classically the singularity corresponds to a delta function of negative mass located at the surface of the bubble.   Presumably smoothing it out quantum mechanically would require some source of negative matter in the theory.  While such a resolved bubble apparently represents an instability in the theory presumably so would the negative matter necessary to resolve it.   There is no such obvious objection in the globally asymptotically flat bubbles and it would be interesting to investigate this case further.
	
\subsection{ Example III}

One can find bubbles with arbitrary initial acceleration simply by tuning the slope of $\gamma$ at the bubble surface appropriately.   However one would like to know whether there are bubbles which expand outwards indefinitely or after some brief expansion they all turn around and collapse.  While this is a question which apparently must be addressed numerically, as mentioned in the previous section one can find the short time behavior of the expansion of the bubble via a perturbation series in time around the initial data surface.   Presumably bubbles which initially expand rapidly are good candidates for bubbles which expand indefinitely.  Examining such small time expansions shows one can make very rapidly expanding bubbles by letting $\beta'(r_0,0)$ be small and positive and $\beta''(r_0,0)$ be of order one and negative.  Of course one must at the same time keep $\beta$ from actually going through a zero, as well as fullfilling the previously mentioned conditions (\ref{condit1}, \ref{condit2}).   One such asymptotically flat example is given by
\be \label{hyper1}
\beta(r,0) = \epsilon \, r_0^2 \Big(1 - \frac{r_0}{r} \Big) - \frac{c_2 \, r_0^2 \Big(1 - \frac{r_0}{r} \Big)^2}{1 + \frac{c_2}{\epsilon} \Big(1 - \frac{r_0}{r} \Big)} + \frac{r^2}{4}  \Big(1 - \frac{r_0^2}{r^2} \Big)^4
\ee
and
\be
\alpha(r,0) = \frac{r^2}{4}
\ee
If one fixes gauge so that
\be
g_{tt} (t,r) = g_{t t} (r) = -1 +\tau_1 \Big(\frac{r}{r_0} - 1\Big) + \mathcal{O}\Big((r - r_0)^2\Big)
\ee
then one finds that to leading order in $\epsilon$ (term by term)
$$
\alpha(r_0,t) = \frac{r_0^2}{4} \Big[1 + \frac{1}{\epsilon} \frac{t^2}{r_0^2}  + \frac{c_2}{3 \epsilon^3 } \frac{ t^4}{ r_0^4}  + \frac{c_2^2}{10 \epsilon^5 } \frac{ t^6}{ r_0^6} + \frac{c_2^3}{35 \epsilon^7} \frac{ t^8}{ r_0^8} + \frac{c_2^4}{126 \epsilon^9} \frac{ t^{10}}{ r_0^{10}}
$$
\be
 + \frac{c_2^5}{462 \epsilon^{11}} \frac{ t^{12}}{ r_0^{12}}  + \frac{c_2^6}{1716 \epsilon^{13}} \frac{ t^{14}}{ r_0^{14}} + \mathcal{O}\Big(\frac{ t^{16}}{ r_0^{16}} \Big) \Big]
\ee
$$
\beta(r,t) = r_0 (r - r_0) \Big[\epsilon - \Big(1 + \frac{\tau_1}{4} \Big) \frac{t^2}{r_0^2}  - \frac{c_2}{3 \epsilon^2 } \frac{ t^4}{ r_0^4} - \frac{c_2^2}{10 \epsilon^4 } \frac{ t^6}{ r_0^6} -\frac{c_2^3}{35 \epsilon^6} \frac{ t^8}{ r_0^8} - \frac{c_2^4}{126 \epsilon^8} \frac{ t^{10}}{ r_0^{10}}
$$
\be
- \frac{c_2^5}{462 \epsilon^{10}} \frac{ t^{12}}{ r_0^{12}}  - \frac{c_2^6}{1716 \epsilon^{12}} \frac{ t^{14}}{ r_0^{14}} + \mathcal{O}\Big(\frac{ t^{16}}{ r_0^{16}} \Big) \Big] + \mathcal{O} ( (r - r_0)^2 )
\ee
$$
W(r,t) = \Big(\frac{r}{r_0} - 1 \Big) \Big[\frac{1}{\epsilon} - \Big(1 + \frac{\tau_1}{4} \Big) \frac{t^2}{\epsilon^2 r_0^2}  + \frac{c_2}{3 \epsilon^4 } \frac{ t^4}{ r_0^4}  + \frac{c_2^2}{10 \epsilon^6 } \frac{ t^6}{ r_0^6} + \frac{c_2^3}{35 \epsilon^8} \frac{ t^8}{ r_0^8} + \frac{c_2^4}{126 \epsilon^10} \frac{ t^{10}}{ r_0^{10}}
$$
\be
 + \frac{c_2^5}{462 \epsilon^{12}} \frac{ t^{12}}{ r_0^{12}}  + \frac{c_2^6}{1716 \epsilon^{14}} \frac{ t^{14}}{ r_0^{14}} + \mathcal{O}\Big(\frac{ t^{16}}{ r_0^{16}} \Big) \Big] + \mathcal{O} ( (r - r_0)^2 )
\ee
Note then if $\epsilon \ll 1$ each term in the series is parametrically larger than that which came before it and so the bubble is expanding much faster than any exponential.  At the same time the value of $\beta$ near the bubble is becoming very small.   If the pattern seen above continued to all orders then the bubble would either reach infinite area in some finite time (i.e. get to null infinity) or $\beta$ would go through a zero and presumably result in a naked singularity.   Of course while writing out the first few terms is suggestive it does not in any sense dictate the form of the higher order terms and the terms listed only necessarily determine the behavior of the given metric functions for times short compared to $\epsilon  r_0$.   Numerically evaluating the first several coefficients seems to indicate that at least the first several coefficients are reasonably approximated by the leading order term in $\epsilon^{-1}$ for $\epsilon \lesssim 1/100$ and $c_2 \sim 1$.  For $c_2 \sim 1$ and $\epsilon \ll 1$ the bubble has mass
\be \label{hypermass1}
M \sim \frac{2 \pi^2}{16 \pi G} 6 r_0^2
\ee
which is to say it is undergoing this rapid expansion despite the fact that it is relatively heavy.

There are many similar variations of (\ref{hyper1}) and in particular if one takes, for example,
\be \label{hyper2}
\beta(r, 0) = \frac{r^2}{4} \Bigg[ \frac{\epsilon (\frac{r_0}{r})^2 \Big( 1 -  (\frac{r_0}{r})^2\Big)}{1 + \frac{c_2}{\epsilon} \Big( 1 -  (\frac{r_0}{r})^2\Big)} + \Bigg( 1 -  \Big(\frac{r_0}{r}\Big)^4\Bigg)^3 \Bigg]
\ee
and $\alpha(r,0) = r^2/4$ one finds a solution with a qualitatively similiar expansion of the bubble at small times and mass nearly a third of (\ref{hypermass1}).   In particular then one can feel confident that (\ref{hyper2}) will not form an event horizon if and unless the bubble first collapses to a fraction of its original size.   One can also write initial data for such rapidly expanding bubbles in the asymptotically AdS case with qualitatively similar features to their asymptotically flat cousins.

 \setcounter{equation}{0}
\section{Stress Tensor}

In the AdS case, one would like to understand the state in the gauge theory dual to these bubbles.  As a first step towards this goal, consider what kind of matter they correspond to by calculating the holographic stress energy tensor \cite{Skenderisetal}.  To compute this quantity one needs the aymptotic metric through $\mathcal{O}(1/r^{4})$.  Solving the Einstein equation to this order for metric (\ref{timedepmet}) taking
\be
\alpha(r,t) = \frac{r^2}{4} \Big(1 + \delta \alpha(t)  \, \frac{l^4}{r^4} +  \mathcal{O}\Big(\frac{1}{r^{4 + \epsilon}} \Big) \Big)
\ee
\be
\beta(r,t) = \frac{r^2}{4} \Big(1 + \delta \beta(t)  \,  \frac{l^4}{r^4} + \mathcal{O}\Big(\frac{1}{r^{4 + \epsilon}} \Big) \Big)
\ee
and
\be
W(r,t) = \frac{r^2}{l^2} + 1 - \delta_{rr}(t) \frac{l^2}{r^2} +  \mathcal{O}\Big(\frac{1}{r^{2 + \epsilon}}\Big)
\ee
one finds the Einstein equations to leading order imply
\be \label{ttleadingorder}
g_{tt} (t,r) = -\Big[\frac{r^2}{l^2} + 1 - \Big(\delta_{r r} + 2 \, \delta \alpha  + \delta \beta \Big) \frac{l^2}{r^2} + \mathcal{O}\Big(\frac{1}{r^{2 + \epsilon}}\Big) \Big]
\ee
Then one finds the holographic stress energy tensor
\be
16 \pi G < T_{\theta \theta} > = \delta_{rr} + \frac{1}{4} + 4 \, \delta \alpha 
\ee 
\be
16 \pi G < T_{\psi \psi} > = \sin^2 \theta \, \Big(\delta_{rr} +\frac{1}{4} +4 \, \delta \alpha \cos^2 \theta + 4 \, \delta \beta \sin^2 \theta \Big)
\ee 
\be
16 \pi G < T_{\phi \phi} > = \cos^2 \theta \, \Big( \delta_{rr}   + \frac{1}{4} +4 \, \delta \alpha \sin^2 \theta + 4 \, \delta \beta \cos^2 \theta \Big)
\ee 
\be
16 \pi G < T_{\psi \phi} > = 4 \sin^2 \theta \, \cos^2 \theta \,  ( \delta \beta - \delta \alpha)
\ee 
and
\be \label{Ttt}
16 \pi G <T_{t t} > = 3 \delta_{r r} + 4 (2 \, \delta \alpha  + \delta \beta ) + \frac{3}{4}
\ee
which, once integrated over the asymptotic $S^3$, simply leads to the energy (\ref{EAdsdef}) aside from a constant offset which may be interpreted as a casimir energy.  Note that the stress tensor is traceless as expected. 
Further if the $S^3$ is squashed (i.e. $\delta \alpha \neq \delta \beta$) the pressures are distinct from those of AdS-Schwarzschild.  The stress tensor $\mathbf{T}$ may be written to make the  $S^2 \times S^1$ symmetry explicit:
\be
 \mathbf{T} = \rho e_t e_t +  p_\alpha \, (\sigma_1 \sigma_1 + \sigma_2 \sigma_2) + p_\beta \, \sigma_3 \sigma_3
\ee
where 
\be \label{S2press}
p_\alpha =\frac{1}{16 \pi G} \Big( \delta\alpha  + \frac{\delta_{rr}}{4} + \frac{1}{16}\Big)= \frac{1}{16 \pi G} \Big( \frac{1}{16} + \frac{2 G M}{3 \pi l^2} + \frac{1}{3} (\delta \alpha - \delta \beta)  \Big)
\ee
is the pressure along the $S^2$ and $M$ is the mass (\ref{EAdsdef}), 
\be \label{S1press}
p_\beta =\frac{1}{16 \pi G} \Big(\delta\beta  + \frac{\delta_{rr}}{4} + \frac{1}{16}\Big) = \frac{1}{16 \pi G} \Big( \frac{1}{16} + \frac{2 G M}{3 \pi l^2}  + \frac{2}{3} (\delta \beta - \delta \alpha) \Big)
\ee
is the pressure along the $S^1$, $\rho = T_{t t}$ as defined in (\ref{Ttt}), and $e_t$ is a unit timelike vector.  Note in AdS $\delta \alpha$ and $\delta \beta$ are generically dynamical quantities and the forms of the pressures suggests the system will become unstable, especially if the magnitude of $\delta \beta - \delta \alpha$ can be made large with respect to the mass and casimir terms.   Making $\delta \beta - \delta \alpha$ more positive, for example, increases the pressure along the $S^1$ and decreases the pressure along the $S^2$ (including possibly making it more negative) and hence tends to increase the size of the perturbation. 

In fact, one can fairly easily produce large negative pressures at a relatively low cost in mass.  Consider, for example, either a bubble or topologically trivial metric containing a squashed $S^3$ where at large r
\be \label{pressuregamma1}
\gamma = 1 + \frac{1}{a + \epsilon \frac{r^4}{l^4}}
\ee
where $\vert a \vert  \gg 1$.  Then initially
\be
\delta \beta - \delta \alpha  = \frac{1}{\epsilon}
\ee
and the magnitude of this term can be made arbitrarily by taking $\epsilon$ to be small.  It is easy to check the mass of such a spacetime for $\epsilon \ll 1$ and for $a \gg 1$  is given by
\be \label{asymM1}
M \sim \frac{l^2}{a \epsilon} \Bigg[ 1 + \mathcal{O}\Bigg( \frac{1}{a}, \sqrt{\frac{\epsilon}{a}} \Bigg) \Bigg]
\ee
and so is parametrically smaller than $\epsilon^{-1}$.   Note in (\ref{asymM1}) I have implicitly assumed any contributions to the mass due to the deviation of $\gamma$ from (\ref{pressuregamma1}) is smaller than the given term.  This may always be taken to be true for topologically trivial solutions, but if one  wishes to consider other $\gamma$ near the origin or include a bubble it will only be so for $a \epsilon$ appropriately small.   
	
	Then one may find arbitrarily large pressures of both signs parametrically larger than the mass,  provided one takes the product of $a$ and $\epsilon$ positive so as to avoid a divergence in $\gamma$.   The stress tensor remains traceless, of course, so in the case where the pressures are much larger than the mass and casimir terms $p_\alpha \sim - p_\beta$; in physical terms if there is a large positive pressure along the $S^1$ there will be a large tension along the $S^2$ and vice versa.  Note the perturbation (\ref{pressuregamma1}) always remains small, although it remains of order $1/a$ over a larger and larger region as one takes $\epsilon \rightarrow 0$.    One might be tempted to take $\vert a \vert$ to be sufficientlly large to make $M$ to go to zero as $\epsilon \rightarrow 0$.  While in the strictly classical sense this would be sensible, quantum mechanically one would find the expectation number of gravitons would become very small and supergravity would no longer be a reliable guide.\footnote{I would like to thank D. Marolf for making this observation.}  If instead one takes $\vert a \vert$ to be large but fixed as $\epsilon \rightarrow 0$ this complication may be avoided.

One may write down a variety of similar initial data to (\ref{pressuregamma1}).  For example, if one takes
\be \label{pressuregamma2}
\gamma = 1 + \frac{1}{a e^{-\frac{\epsilon}{a}  \frac{r^4}{l^4}}  + \epsilon \frac{r^4}{l^4}}
\ee
again one finds that
\be
M \sim \frac{l^2}{a \epsilon} \Bigg[ 1 + \mathcal{O}\Bigg( \frac{1}{a}, \sqrt{\frac{\epsilon}{a}} \Bigg) \Bigg]
\ee
with, as above, the same minor caveat discussed below (\ref{asymM1}).  As explained in the next section, it will be considerably easier, however, to understand certain aspects of the evolution of (\ref{pressuregamma2}) than of (\ref{pressuregamma1}).  Note the precise form of the exponential has been chosen so it is significantly different from one only in a region where $\gamma$ falls off as $r^{-4}$.   Hence $ \vert 1 - \gamma \vert \lesssim 1/{\vert a \vert}$.

Then if $\vert a \vert \gg 1$ and $\vert \epsilon \vert \ll 1$, initial data of the types described violate the strong, weak, and dominant energy conditions parametrically.  Hence boosted observers (even those moving at low velocities) will see regions of negative energy.  Note the positive energy theorems \cite{PosEnergyAdS} do not protect one against such phenomena as the total energy on any given Cauchy slice (defined via the global timelike killing vector) will remain positive.  It is well known that quantum mechanically it is possible to violate the classical energy conditions for a sufficiently short period of time (see e.g. \cite{FordRoman} and references therein).  In such a case these violations are only transient behavior instead of a signal of instability.  Hence it is important to understand how long such violations may last and if in fact they become worse as time goes on.  This is equivalent to asking a question regarding the evolution of such data in AdS, the subject of the next section.

 \setcounter{equation}{0}
\section{AdS Evolution}

As shown in the last section, one can write down initial data for both  bubbles and topologically trivial metrics which violates the classical energy conditions.   To understand whether the quantum inequalities are violated one must get at least some bounds on the time evolution.  For short times, one can consider in the full nonlinear theory a perturbation series in time about the initial data slice as described earlier and finds evidence that at least some such perturbations tend to grow in time.   Since the relevant perturbation is small, one may also study the time evolution using the linearized equations.   There one finds that not only can perturbations violating the energy bounds last for a time of order $l$ but in fact they can grow significantly within such a time period.  Given this apparent instability in the dual field theory, it is natural to inquire as to whether there is a dual gravitational instability.  Finally I discuss why the standard mode sum analysis does not necessarily prevent the growth of these perturbations as one might have expected.

\subsection{Nonlinear Results}

It is handy to fix a gauge such that $\alpha(r, 0) = r^2/4$ and $g_{tt}(r,t) = g_{tt}(r)$.   Requiring that the asymptotic metric take the canonical form
$$
ds^2 = -\Big(\frac{r^2}{l^2} + 1 + \mathcal{O}\Big(\frac{1}{r^2} \Big) \Big)dt^2 + \frac{dr^2}{\frac{r^2}{l^2} + 1 + \mathcal{O}\Big(\frac{1}{r^2} \Big)}
$$
\be
 + (\frac{r^2}{4} + \mathcal{O}\Big(\frac{1}{r^2}\Big)\Big) \Big(d \bar{\theta}^2 + \sin^2(\bar{\theta}) d\bar{\phi}^2 \Big) + (\frac{r^2}{4} + \mathcal{O}\Big(\frac{1}{r^2}\Big)\Big) \Big({d \bar{\psi} + \cos(\bar{\theta}) d\bar{\phi}}\Big)^2
\ee
at each order in $t$ is sufficient to fix $g_{tt} (r)$, at least through the terms considered below.   There does not appear to be any simple argument regarding the asymptotics of the metric to all orders so the present discussion will be limited to considering the first several terms;  in particular the results  of this subsection hold through order $(t/l)^{12}$.  

Consider an expansion of the form
\be \label{gammaasym}
\gamma = 1 + \Sigma_{n = 4}^{\infty} b_n \frac{l^n}{r^n} + \ldots
\ee
where the omitted terms may include exponentially damped terms and other terms smaller than any power of $r^{-1}$ which one will not need to keep track of explicitly.   Defining $\delta(t)$ as the leading asymptotic correction to $\beta(r,t)$, i.e.
\be
\beta(r,t) = \frac{r^2}{4} + \delta(t) \frac{L^4}{r^2} + \mathcal{O}\Big(\frac{L^4}{r^4} \Big)
\ee
one finds
$$
\delta(t)  = b_6 \frac{t^2}{l^2} + \frac{21 b_6 + 24 b_8 + 4 b_4 \Big(b_4 - \frac{M_0}{l^2} \Big)}{9} \frac{t^4}{l^4} 
$$
\be \label{delta1}
 + \frac{ 2 \Big[120 b_{10} + 200 b_8 + b_6 \Big(80 b_4 + 81 -36 \frac{M_0}{l^2} \Big) + 20 b_4 \Big(b_4 - \frac{M_0}{l^2} \Big) \Big]}{45} \frac{t^6}{l^6} +  \mathcal{O}\Big(\frac{t^8}{l^8} \Big)
\ee
where $M_0$ is proportional to the mass M (\ref{EAdsdef}) of the solution:
\be
M_0 = \frac{16 \pi G}{2 \pi^2} M
\ee
Further
\be
\alpha (r,t) = \frac{r^2}{4} - \frac{\delta(t)}{2} \frac{L^2}{r^2} + \mathcal{O}\Big(\frac{L^4}{r^4} \Big)
\ee
and
$$
W(r,t) = \frac{r^2}{l^2} + 1 +\frac{\Big(4 b_4 - \frac{M_0}{l^2}\Big) l^2}{3 r^2} + \frac{(b_4 + 2 b_6) l^4}{r^4}
$$
$$
+  \frac{l^6}{r^6} \Bigg[\frac{2 b_4 \Big(b_4 - \frac{M_0}{l^2} \Big) + 15 b_6 + 24 b_8}{9}+  \frac{8 b_4 b_6}{3} \frac{t^2}{l^2} 
$$
\be
+ \frac{8}{27} \Big[4 b_4^2 \Big( b_4 - \frac{M_0}{l^2} \Big) + 3 b_4 (7 b_6 + 8 b_8) + 27 b_6^2\Big] \frac{t^4}{l^4} +   \mathcal{O}\Big(\frac{t^6}{l^6} \Big) \Bigg] +  \mathcal{O}\Big(\frac{l^8}{r^8} \Big)
\ee
Noting that time dependence does not occur in $W(r,t)$ until several orders beyond the leading nontrivial term, energy conservation (\ref{EAdsdef}) dictates the simple connection between the leading order corrections to $\alpha(r,t)$ and $\beta(r,t)$.  At least to leading order in $r^{-1}$ this result is fairly easy to derive for all times.  If one writes
\be
g_{t t}(r) = -\frac{r^2}{l^2} - 1 + t_0 \frac{l^2}{r^2} + \mathcal{O}\Big(\frac{1}{r^{2 + \delta}} \Big)
\ee
for some $\delta > 0$ then from the equations of motion at leading order  in $r^{-1}$ (\ref{ttleadingorder}) and the fact that energy  (\ref{EAdsdef}) will be conserved under evolution one finds that
\be
\delta_{r r} = 4 t_0 - \frac{M_0}{l^2}
\ee
and
\be
\delta \beta + 2 \delta \alpha = \frac{M_0}{l^2} - 3 t_0
\ee
Thus if $\delta \beta$ grows as a function of time $\delta \alpha$ contracts and vice versa.   Note this is precisely what one would expect from the results involving pressures in the dual gauge theory from the previous section.    It also implies that one can think of the asymptotic dynamics entirely in terms of the squashing of a $S^3$.  Finally note that while conservation of energy restricts the form of the asymptotic metric evolution it does not place any limitations on the magnitude of the squashing.

While it would be interesting to better understand the generic time dependent asymptotics, consider, as an example, cases such as (\ref{pressuregamma2}) where the only nonzero polynomial expansion coefficient is $b_4 = 1/\epsilon$.  Then
$$
\delta(t) = \frac{4 b_4 \Big(b_4 - \frac{M_0}{l^2} \Big)}{9} \frac{t^4}{l^4} +  \frac{8 b_4 \Big(b_4 - \frac{M_0}{l^2} \Big)}{9} \frac{t^6}{l^6} 
$$
$$
+  \frac{4 b_4 \Big[b_4 \Big(80 b_4 - 156 \frac{M_0}{l^2} \Big) + 281\Big(b_4  - \frac{M_0}{l^2} \Big) + 76 \frac{M_0^2}{l^4}\Big] }{945} \frac{t^8}{l^8} 
$$
$$
+  \frac{8 b_4 \Big[b_4 \Big(1216 b_4 - 2256 \frac{M_0}{l^2} \Big) + 1363\Big(b_4  - \frac{M_0}{l^2} \Big) + 1040 \frac{M_0^2}{l^4}\Big] }{8505} \frac{t^{10}}{l^{10}} 
$$
$$
+  \frac{8 b_4}{467775} \Bigg[b_4^2 \Big(7680 b_4 - 27840 \frac{M_0}{l^2} \Big) + b_4 \Big(133520 b_4  - 235764 \frac{M_0}{l^2} \Big)
$$
\be \label{deltab4}
 + (29232 b_4 - 9072  \frac{M_0}{l^2} \Big) \frac{M_0^2}{l^4} + 69987 \Big(b_4  - \frac{M_0}{l^2} \Big)+ 102244 \frac{M_0^2}{l^4}\Bigg] \frac{t^{12}}{l^{12}} +  \mathcal{O}\Big(\frac{t^{14}}{l^{14}} \Big)
\ee
Note that,at least through the order studied, $\delta(t)$ increases more and more rapidly  if $b_4 \gg M_0/l^2$ and $b_4 \gg1$ and in particular does so parametrically faster than an exponential.   In fact one can check that even for a mass as large as $b_4 l^2/10$ all the terms studied are positive and if $b_4 \gg 1$ increasing parametrically.   This matches nicely with the expectations from the stress tensor discussed earlier.   It is important, however, to note that since (\ref{deltab4}) involves more powers of $b_4$ at higher orders one only trusts the expansion for $t \lesssim 1/b_4$ if $b_4 \gg 1$ and for a rather shorter time if $b_4 \sim 1$.   Furthermore in such a time $\delta$ will not change very much.  The form of the terms is suggestive but without any argument restricting the form of terms to all orders in $t$ one can not make any definite conclusions regarding the long time behavior.

\subsection{Linearized Evolution}
Given a small initial perturbation, the linearized equations should be a reliable guide for the evolution at least if and until such a perturbation evolves in such a way that it grows dramatically.  Consider the  linearized metric
$$
ds^2 = \Bigg[-\Big(\frac{r^2}{l^2} + 1 \Big) + \delta g_{tt}(r,t)\Bigg] dt^2 + \Bigg[\frac{1}{\frac{r^2}{l^2} + 1 } + \delta g_{rr}(r,t)\Bigg] dr^2
$$
\be \label{linmetric}
  +\Big[\frac{r^2}{4} + \delta g_{\bar{\theta} \bar{\theta}} (r,t)\Big] \Big(d \bar{\theta}^2 + \sin^2(\bar{\theta}) d\bar{\phi}^2 \Big)+\Big[\frac{r^2}{4} + \delta g_{\bar{\psi} \bar{\psi}} (r,t)\Big]\Big({d \bar{\psi} + \cos(\bar{\theta}) d\bar{\phi}}\Big)^2
\ee
It is furthermore useful to define
\be
\delta g_{\bar{\psi} \bar{\psi}} (r,t) = C_0(r,t) \frac{l^2}{r^2} +  \delta g_{\bar{\theta} \bar{\theta}} (r,t)
\ee
Then the linearized Einstein equations yield
$$
\delta g_{rr} = \frac{1}{3 r^4 (l^2 + r^2)^2} \Big[ 3 r^2 K_0 - 4 l^4 ( 3 l^2 + 4 r^2) C_0 + 4 l^4 ( l^2 + r^2) r C_0' 
$$
\be \label{dgrr}
 - 12 l^2 r^2 (l^2 + 2 r^2)  \delta g_{\bar{\theta} \bar{\theta}} +12 l^2 r^3  (l^2 + r^2)  \delta g_{\bar{\theta} \bar{\theta}}' \Big]
\ee
where $K_0$ is an integration constant which will vanish if one requires the perturbations to be regular near the origin
$$
 \delta \ddot{g}_{\bar{\theta} \bar{\theta}} = -\frac{1}{12 l^6 r^3} \Big[6 r (l^2 + 2 r^2) K_0 - 8 l^4 r (l^2 + 2 r^2) C_0  - 4 l^4 (5 l^4 + 7 l^2 r^2 + 2 r^4) C_0'  
 $$
$$
 +4 l^4 r (l^2 + r^2)^2 C_0'' - 6 l^4 r^5 \delta g_{t t} + 3 l^4 r^4 (l^2 + r^2)  \delta g_{t t} '
 $$
 \be
  - 24 l^2 r^5 \delta g_{\bar{\theta} \bar{\theta}} + 12 l^2 r^4 (l^2 + r^2) \delta g_{\bar{\theta} \bar{\theta}}' \Big]
\ee
 and
 \be \label{linflateqn}
  l^2 \ddot{C}_{0} =\Big(1 + \frac{r^2}{l^2} \Big) \Big[(l^2 + r^2) C_0'' - (5 l^2+ 3 r^2) \frac{C_0'}{r} \Big]
   \ee
 where primes denote derivatives with respect to r and dots derivatives with respect to t.   Regular perturbations correspond to those in which asymptotically $\delta g_{r r}$ is of order $r^{-6}$, $\delta g_{\bar{\theta} \bar{\theta}}$, and $\delta g_{t t}$ are of order $r^{-2}$, and $C_0$ goes to a finite value, while near the origin $\delta g_{r r}$ and $\delta g_{t t}$ are of order $r^2$, $\delta g_{\bar{\theta} \bar{\theta}}$ is of order $r^4$, and $C_0$ is of order $r^6$.  Note that one may regard $C_0 (r,t)$ as the squashing of the sphere and via (\ref{linflateqn}) solve for it independently of the other metric perturbations.  The other perturbations will be sourced by $C_0$ and one may use the above equations to determine them.  While the results are typically complicated, it is worth noting there is one case in which $\delta g_{r r}$ vanishes.   If one takes
 \be
 \delta g_{\theta \theta} = -\frac{C_0}{3} \frac{l^2}{r^2}
 \ee
 and requires the perturbations to be regular then $\delta g_{r r} = \delta g_{t t} = 0$. More generically, if one takes a gauge where $\delta g _{t t} (r)$ then $\delta g_{r r}$ is static at leading order.  Further if one expands (\ref{dgrr})-(\ref{linflateqn}) in a power series in terms of $r^{-1}$ one finds that $\delta g_{r r}$ remains static at least through the first dozen nontrivial terms in the linearized theory.  
 
One may solve (\ref{linflateqn}) via a mode sum but, as discussed later, the behavior of the perturbation at large distances is rather opaque.  For the moment consider specifying initial data and analyzing the asymptotics of the resulting spacetime.  For the sake of simplicity I will restrict the present discussion to time symmetric initial data.  Further consider spacetimes where one may then expand $C_0$ asymptotically as
\be \label{C0exp}
C_0 (r,t) = \Sigma_{m = 0}^{m  = \infty} b_m (t)  \frac{l^m}{r^m} + \ldots
\ee
for at least some finite period of time $t_1$ where the omitted terms may include exponentially damped and other terms smaller than any order in $r^{-1}$ which I will not write explicitly.  In particular suppose one takes initial data where $\gamma(r)$ is a sum of rational functions of polynomials and exponentials for $r \geq r_1$ for some constant $r_1$ with the previously described asymptotics (i.e. (\ref{gammaasym}) ).   Then considering (\ref{linflateqn}) for a few moments (or if one wishes the time discretized version of (\ref{linflateqn}) ) should convince the reader that, at least until the time $t_1$ it takes signals to propagate from $r = r_1$ to near the boundary, the spacetime will necessarily have an asymptotic expansion of the form (\ref{C0exp}). Then (\ref{linflateqn}) implies the coefficients $b_m$ vanish for odd m while for even m
\be
l^2 \ddot{b}_0 = 12 b_2
\ee
and
\be \label{linexp5}
l^2 \ddot{b}_{m+2} = (m+4) (m+8) b_{m+4} + 2 (m+2) (m+7) b_{m+2} + m (m+6) b_m
\ee
for $m \geq 0$.   Expanding the $b_m$ as a functions of time
\be
b_m = \Sigma_{n = 0}^{n = \infty} c_n^{(m)} \frac{t^{2n}}{l^{2n}} 
\ee
one finds 
 $$
c_{n+1}^{(m + 2)} = \frac{1}{(2n + 2) (2n + 1)} \Big[ (m+4)(m+8) c_n^{(m+4)} 
$$
\be  \label{linexp6}
+ 2 (m+2) (m+7) c_n^{(m+2)} + m (m+6) c_n^{(m)} \Big]
\ee
where for the sake of compactness I have defined $c_n^{(-m)} = 0$.   Note if at any order $n_c$ in $t$ all the coefficients $c_{n_c}^{m}$ have a single sign then necessarily  all terms $c_n^{m}$ for $n > n_c$ also posses that same sign.  In particular if one considers initial data where all the coefficients of inverse powers of r are positive then the asymptotic value of the perturbation increases monotonically at least for $t < t_1$.

For the sake of completeness, I should mention there are several other tacts one might use to derive (\ref{linexp6}).  One is simply to observe the mode solutions, which will be reviewed later in this section, can be written in terms of polynomials of $t$ (via expanding $\cos(\omega t)$) and, at least for $r > l$, of $r^{-2}$.  To make a precise argument along these lines, however, seems to be somewhat nontrivial.  One may also expand the initial data first in terms of time and the expand each of the resulting coefficients at large r.  However, one might worry in this case that the expansion as a function of time might fail to converge at some time rather smaller than $t_1$ due to nontrivial evolution in the interior of the spacetime.

Turning now to specific examples, suppose one considers time symmetric initial data of the form of (\ref{pressuregamma2})
\be \label{pressuregamma3}
\gamma = 1 + \frac{1}{a e^{-\frac{\epsilon}{a}  \frac{r^4}{l^4}}  + \epsilon \frac{r^4}{l^4}}
\ee
for $r \geq r_1$ and is smoothly, although not necessarily analytically, matched onto some regular solution for $r \leq r_1$.  One may use the above expansion (\ref{linexp6}) to find the asymptotic time behavior until signals from the interior region may propagate to the asymptotic region.  For the particular data under consideration (\ref{pressuregamma3}) provided one takes $r_1 \ll l$ this will result in a constant pressure of arbitrary magnitude for times on order of $\pi l/2$.  Hence one may parametrically violate the quantum energy inequalities.   As seen in the previous section these strictly static pressures will become dynamical in the non-linear theory, although such modifications are necessarily small corrections.

More generic initial data will result in a dynamical pressure even in the linearized theory.  In particular, consider time symmetric initial data of the form
\be \label{pressuregamma4}
\gamma = 1 + \frac{1}{a e^{-\frac{\epsilon}{a}  \frac{r^{n_{0}}}{l^{n_{0}}}}  + \epsilon \frac{r^{n_{0}}}{l^{n_{0}}}}
\ee
for $r \geq r_1$ where $n_0 \geq 4$.  Again smoothly match this data onto a regular solution for $r \leq r_1$ where $r_1 \ll l$.   In this case one may verify that the mass of the solution will scale as
\be \label{pressuregamma5}
M \sim \frac{l^2}{a^{2 - \frac{4}{n_0}} \epsilon^{\frac{4}{n_0}} G} \Bigg[ 1 + \mathcal{O}\Bigg( \frac{1}{a}, \Big(\frac{\epsilon}{a} \Big)^{\frac{2}{n_0}} \Bigg) \Bigg]
\ee
again with the same minor caveat discussed below  (\ref{asymM1}) if one has a bubble or other nontrivial structure in the interior of the spacetime.  Since there is only a single coefficient in the expansion of $r^{-n}$ of (\ref{pressuregamma4}) the perturbation will be monotonically increasing asymptotically, at least until a time of order $\pi l/2$.  In particular if one takes $n_0 = 6$ the asymptotic value of $C_0$ to the first several orders in time is
\be \label{C0expan1}
b_0 = \frac{l^2}{4 \epsilon} \Big[ 6 \frac{t^2}{l^2} + 14 \frac{t^4}{l^4} + \frac{108}{5} \frac{t^6}{l^6} + \frac{2734}{105} \frac{t^6}{l^6} + \frac{125444}{4725} \frac{t^8}{l^8} + \frac{125236}{51975} \frac{t^{10}}{l^{10}} + \mathcal{O} \Big(\frac{t^{12}}{l^{12}} \Big) \Big]
\ee
While one knows that all the terms in the expansion have the same sign working out the combinatorics given by (\ref{linexp6}) is a nontrivial problem even in this simple case.  To get some sense of the results one can focus on the first two dozen terms.   The coefficients eventually decrease from those given in (\ref{C0expan1}) reaching order one by $n = 14$.   At least for the successive ten terms the numerical coefficient of each term is roughly half the one before it.   Hence, at least through the considered terms, $b_0$ increases in magnitude much more rapidly than an exponential.   In fact, if the described pattern held to all orders one would expect the series to diverge in time $\sim \sqrt{2} l$.    One can at least definitely place some lower bounds on the magnitude of $a_0$.   By time $t=l$ 
\be
b_0 \epsilon  \gtrsim 46
 \ee
by time $t/l = \pi/2 - 1/10$ 
\be \label{b02}
b_0 \epsilon  \gtrsim 5 \times10^5
 \ee
 and by time $t/l = \pi/2 - 1/100$
 \be \label{b03}
b_0 \epsilon  \gtrsim 5 \times10^6
\ee
This, of course, does not constitute a proof of an instability in linearized AdS.  If the perturbations remain finite, however, it is difficult to understand the origin of such large numbers (\ref{b02}, \ref{b03}).

Turn now to the usual solution of the linearized problem via mode solutions.  Defining
\be
y = \frac{r^2}{r^2 + l^2}
\ee
and taking $C_0 = e^{i \omega t} f(y)$ one finds the only mode solutions which are regular asymptotically and meet the boundary conditions near the origin are
\be \label{modsol}
f(y) = y^3 \, \, {}_{2} F_{1} \, (3 - \frac{l \omega}{2},3 + \frac{l \omega}{2},4,y)
\ee
where
\be
\omega = \frac{2 m}{l}
\ee
and m is a natural number greater than or equal to three.   The hypergeometric functions may expressed in terms of the complete Jacobi polynomials
\be
P_n^{(2,3)} (x) = \frac{(-1)^n (n+3) (n+2) (n+1)}{6} \, \, {}_{2} F_{1} \, (3 - \frac{l \omega}{2},3 + \frac{l \omega}{2},4,y)
\ee
where $n = m - 3$ is a nonnegative integer and $x = 2 y - 1$.   The weight for these polynomials is given by $(1-x)^2 (1+x)^3$ and they satisfy an orthogonality relation
\be
\int_{-1}^{1} dx (1- x)^2 (1+x)^3 P_n^{(2,3)} (x)  P_m^{(2,3)} (x) = \frac{ 2^5 (n+2) (n+1)}{(n+5) (n+4) (n+3)} \delta_{m n}
\ee

One might be inclined to conclude from the above observations that the linearized theory is stable.  However, one encounters a subtlety once one considers actually adding modes together to form initial data.  Expanding a function $f(x,t)$ in terms of these modes
\be \label{modesum1}
f(x,t) = \Sigma_{n = 0}^{\infty} a_n e^{\frac{2 i n t}{l}} (1+x)^3 P_n^{(2,3)} (x) 
\ee
where
\be
a_n =  \frac{(n+5) (n+4) (n+3)}{32 (n+2) (n+1)} \int_{-1}^{1} dx f(x,0) (1-x)^2 P_n^{(2,3)} (x) 
\ee
Aside from initial data which is given by a finite series of Jacobi polynomials (a set of measure zero) the resulting series does not converge pointwise at the points $x = \pm 1$ but only in a $L^2$ sense.   In fact at these points the magnitude of each individual term is of order $n^{\frac{5}{2}}$ for large n.   This is not a problem of principle, for in a $L^2$ sense the series will converge to a smooth function, but it does make (\ref{modesum1}) practically useless for finding the leading order time dependence of the asymptotic metric.

More importantly, a general $L^2$ normalizable function which may be written as a sum of these modes (each of which obeys the usual boundary conditions) will not respect the boundary conditions one normally requires for a perturbation to be regular and asymptotically AdS.   Near the origin  ($x = -1$) $L^2$ normalizability  requires merely that
\be \label{norm1}
C_0 \sim \frac{1}{(1 + x)^{2 - \delta}} \sim \Big( \frac{l}{r} \Big) ^{4 - 2 \delta}
\ee
for some $\delta > 0$ or, equivalently, allows a perturbation which diverges as the origin as long as it does not do so as fast as $r^{-6}$.
Near $x = 1$ $L^2$ normalizability only requires that 
\be \label{norm2}
C_0 \sim \frac{1}{(1-x)^{\frac{3}{2} - \delta}} \sim \Big( \frac{r}{l} \Big) ^{3 -2\delta}
\ee
or in other words only requires that the perturbation does not asymptotically grow as fast as $r$.  Note via completeness one may write any such perturbation in terms of normalizable modes.

In fact, there seems to be good reason to believe that this particular type of perturbation with slower than the usual asymptotic falloff conditions may be physically allowed.  That is, it may be sensible (and perhaps even necessary) to weaken the usual boundary conditions slightly for these particular squashing perturbations.  Generically sums of normalizable modes should be allowed unless the sum has a pathology.  Note that, via (\ref{dgrr}), if one chooses a gauge in which $\delta g_{r r}$ obeys the usual falloff conditons (i.e. $\delta g_{r r} = \mathcal{O}(1/r^6) )$ then
\be
\delta \beta + 2 \delta \alpha = \mathcal{O}(r^0)
\ee
This will then be sufficient to ensure that the equation for the energy (\ref{engeneral}) will be finite.  More generically, requiring that the variation of the Hamiltonian will be well defined requires the slightly stronger condition that the perturbations vanish asymptotically.  Note this is still much stronger than the usual requirement that the perturbations fall off as $r^{-2}$.   

Note further that while such variations are larger than usually allowed they will still always be small asymptotically.  Regarding the linearized equations of motion, one merely needs to note that the evolution will simply be given by phase factor evolution in the mode sum and the $L^2$ falloff conditions not be violated under evolution.  In other words, initial data which is $L^2$ normalizable will not evolve into a state which is not $L^2$ normalizable.   On general grounds in the asymptotic region the linearized equations of motion should reliably give the leading order effect of perturbation.   One may also check, for example, that perturbing around the initial data to order $t^2$ the nonlinear effects are smaller than $\mathcal{O}(r^{-4})$ if one requires the variation of the Hamiltonian to be well defined.  Hence one may feel confident that any boundary conditions regarding the metric on the boundary or the extrinsic curvature constructed from the boundary will be preserved under evolution.

While it is difficult to demonstrate explicitly, then, at least at present, there does not seem be any reason to exclude such perturbations and further to believe that perturbations satisfying the usual falloff conditions may not evolve into these more generic perturbations.   In fact, the rapid growth of $C_0$ discussed above, and even its divergence in finite time, may simply be such an evolution.   While such a divergence would seem disastrous for the corresponding CFT, the gravitational description might still be perfectly well defined.   Such a relaxation into the more generic boundary conditions might reasonably then be characterized as natural and expected rather than indicating an instability.

Finally, note while it was convenient above to consider examples with damped exponentials it is not at all clear this feature is present in all or even most perturbations which lead to a growing distortion of the $S^3$.   It would be very interesting to obtain a better understand of such perturbations generically, both in five and higher dimensions.  Further, one would like to find some better method of understand the evolution of the perturbations to be able to discuss their character definitively.

 \setcounter{equation}{0}
\section{Asymptotically Flat Evolution}

In contrast to the AdS case, for asymptotically flat space there does not appear to be any such dangerous growth in squashing perturbations.  Consider linearized perturbations in this case.  One can find a very similar set of evolution equations for the perturbations as in the AdS case.  In this case defining
\be
C_1(r,t) = \delta_{\bar{\psi} \bar{\psi}} (r,t) - \delta_{\bar{\theta} \bar{\theta}} (r,t)
\ee
using the obvious linearized metric (i.e. (\ref{linmetric}) with $l \rightarrow \infty$).
The evolution equation for $C_1$ is
\be \label{linflatpert}
\ddot{C}_1 = C_1'' - \frac{C_1'}{r} - \frac{8 C_1}{r^2}
\ee
where dots denote derivatives with respect to $t$ and primes derivatives with respect to r.  Defining
\be \label{linrescale}
C_1 (r,t) = \sqrt{r} C_2 (r,t)
\ee
(\ref{linflatpert}) becomes simply
\be \label{flatperev}
-\ddot{C}_2 + C_2'' =  \frac{35 C_2}{4 r^2}
\ee
which is to say the evolution equation for a scalar field with a position dependent mass.  This mass is always positive but goes to zero asymptotically.  This will drive any perturbation to infinity at which point  (\ref{flatperev}) will simply act as a massless wave equation.   One might be concerned that if $C_2$ ever asymptotically became nonzero then this would, due to the rescaling (\ref{linrescale}), result in a divergent $C_1$.  This however simply reflects the familiar fact that radiation fields may fall off more slowly near null infinity than the static parts of the metric near spatial infinity.   Nonlinear asymptotically flat perturbations show very similar behavior to these linearized perturbations and there does not appear to be any need for concern in the asymptotically flat case.

 \setcounter{equation}{0}
\section{Discussion}

I have presented a wide variety of single bubbles and topologically metrics containing squashed $S^3$'s in five dimensions.  Exploring such solutions can begin to understand the new and unexpected features of gravity in higher dimensions which have begun to emerge in recent years.  It seems quite likely the ansatz explored extends to higher odd dimensions.   The AdS solutions can also be lifted to higher dimensions via the usual Freund-Rubin compactification.  One would further like to consider adding matter, and in particular, charge to such solutions.

If any of the bubbles described here expand forever the consequences are likely to be profound.   Classically such bubbles would seem perfectly valid initial data which threaten to cutoff null infinity and hence violate cosmic censorship.  One might argue that bubbles in AdS are more likely to stop expanding at some point, either based on the local negative energy of AdS or simply the generic tendency of AdS to make things collapse.  Both arguments would naively, however, apply in locally AdS space as well and in that case Kaluza-Klein bubbles are known to expand forever \cite{VR}.  Furthermore the examples of the static AdS bubbles discussed indicates any such argument must be rather sensitive to the global asymptotics.  For both the asymptotically flat and AdS spaces it is difficult to know which if any bubbles described here would expand forever.   Even if it turned out that numerically exploring several cases determined that expanding bubbles always turned around and collapsed one would like a generic argument for the stability of higher dimensions rather than presuming any counterexample would be immediately obvious.

Quantum mechanically these solutions are also quite interesting.  In the absence of any mechanism to forbid it, one would expect these bubbles to be nucleated quantum mechanically (if nothing else at an order of magnitude larger than the string scale).   If this expansion could ever continue for a long period of time to produce relatively large bubbles it seems difficult to avoid considering this process an instability.   The collapse of light bubbles could also be very interesting quantum mechanically; if one could ever get down to a string scale size bubble before forming an event horizon one would have found a quantum version of cosmic censorship violation.

Perhaps the most interesting open questions related to these solutions lie in AdS/CFT.   If any bubbles do get out to infinity in AdS this would seem to pose a severe challenge for the correspondence.   The gauge theory evolves in a perfectly regular fashion and it is hard to see what could go wrong as the bubble hits the boundary.   On the other hand, even if they do not expand forever states which describe the disappearance of space and its reemergence clearly seem worthy of investigation.

I have pointed out that AdS-CFT implies that the dual guage theory necessarily contains matter which violates all classical and quantum energy bounds.  At least on first appearance this sounds like an instability in the gauge theory.  It is important, however, to consider in detail whether one can make sense of such a field theory.   I have pointed out that perturbations in the dual gravitational description may grow much larger than any scale in the problem would suggest.  It is possible that such growth indicates a gravitational instability or perhaps, as suggested above, the development of asymptotics slightly more generic than that usually allowed.   It is both interesting and important to understand definitively the fate of AdS under such perturbations.

\vskip 1cm
\centerline{\bf Acknowledgments}
\vskip .5cm

It is a pleasure to thank  S.Ross, A. Peet, T. Wiseman, J. Gauntlett, D. Marolf and especially G. T. Horowitz for useful discussions.   This work was supported by grant NSF-PHY-0555669.

 \end{document}